\documentclass[11pt]{article}

\usepackage{amssymb}
\usepackage{euscript}

\newcommand{\B}{\mathfrak{B}}

\newcommand{\Hi}{\mathcal{H}}

\newcommand{\Tr}{\mathrm{Tr}}

\newcommand{\ket}[1]{\lv #1 \ra}
\newcommand{\bra}[1]{\la #1 \rv}
\newcommand{\braket}[2]{\la #1 \vert #2 \ra}

\newtheorem{Theorem}{{Theorem} }

\newtheorem{Lemma}{{Lemma}}

\def\bra#1{\mathinner{\langle{#1}|}}
\def\ket#1{\mathinner{|{#1}\rangle}}
\def\braket#1{\mathinner{\langle{#1}\rangle}}

\title{From Bell Inequalities to Tsirelson's Theorem: A Survey}

\author{
David Avis\thanks{
The author is with The School of Computer Science and GERAD, McGill University,
3480 University, Montr\'eal, Qu\'ebec, Canada H3A 2A7.
}\\
  avis@cs.mcgill.ca
\and
Sonoko Moriyama\thanks{
The author is with Institute for Nano Quantum information Electronics,
University of Tokyo, 7-3-1, Hongo, Bunkyo-ku, Tokyo, 113-8656, Japan.
}\\
  moriso@is.s.u-tokyo.ac.jp
\and
Masaki Owari\thanks{
The author is with Optics Section Blacket Laboratory, Department of Physics, 
and Quantum Information Programme, Imperial College, London, 53 Prince's Gate, South Kensington, London, United Kingdom, SW7 2PG,
and Institute for Nano Quantum information Electronics,
University of Tokyo, 7-3-1, Hongo, Bunkyo-ku, Tokyo, 113-8656, Japan.
}\\
  m.owari@imperial.ac.uk
}

\begin{document}
\maketitle

\begin{abstract}
The first part of this paper contains an 
introduction to Bell inequalities and 
Tsirelson's theorem for the non-specialist.
The next part gives an explicit optimum construction for the
``hard'' part of Tsirelson's theorem.
In the final part we describe how upper bounds on the maximal
quantum
violation of Bell inequalities can be obtained by an extension of Tsirelson's theorem,
and survey very recent results on how exact bounds may be obtained
by solving an infinite series of semidefinite programs.
\end{abstract}

\section{Introduction}
Correlations between observed measurements are a fundamental
resource in quantum information.
In his seminal 1964 paper, Bell \cite{Be64} demonstrated an
inequality that must be satisfied by correlations obtained
classically, but may be violated by a quantum correlation experiment.
Subsequently, the term {\em Bell inequalities} 
has come to describe the set of inequalities that 
characterize correlations
that can be obtained between events that can be well described by
classical physics.
Since the 1980s, a series of experimental results have been published
that apparently demonstrate that quantum correlations do in fact
violate Bell inequalities \cite{experimental papers}.
This raises the question of whether a ``good'' mathematical
characterization of quantum correlation vectors can be obtained.

The first part of the paper is intended for the non-specialist.
We begin by giving a survey of classical correlations, showing
that they can be characterized using
the well studied subject of $L^1$-embeddings.
Next we consider quantum correlations
and give a statement of Tsirelson's theorem \cite{Ci80a}, which gives
a partial characterization for the 2-party case. 
Interestingly, as we show, this characterization can be restated in terms of
$L^2$-embeddings.
Tsirelson's theorem has found many applications in quantum information,
but a direct constructive proof of his theorem is not readily available.
Although this theorem is powerful, even in the two-party case
it does not solve the characterization problem completely.
We discuss this, and give an additional
necessary condition based on no-signalling.
In many applications of Tsirelson's theorem it is necessary
to construct a quantum realization for a given correlation vector.
An explicit construction is not given in his paper, but we give one
here.
We begin by giving  a
simple construction for the two dimensional case of Tsirelson's
theorem, which uses a 
two qubit state. This includes introducing many of the ideas needed
for the general construction. 

In the second part of the paper, we give a new general construction,
that is, states and operators which can represent all 
correlation functions. 
Our method is related to the theory of stabilizer states. 
We also demonstrate the optimality of this result, and give a scheme
for a low dimensional approximation of an optimal representation.

In the third part of the paper we discuss the problem of finding the maximum
violation of a Bell inequality. We see that semidefinite programming (SDP) plays a crucial
role in this optimization problem, and relate it to the results given earlier in the paper.
Finally we review some very recent work involving an infinite hierarchy of SDPs that
gives a non-polynomial time method of characterizing Bell inequalities and
finding maximum violations.


%
%
\section{Classical correlations}\label{class}
Let  $A_1 , ... , A_n$ be a collection of $n$ $0/1$ valued
random variables that belong to a common joint
probability distribution. For $1 \leq i < j \leq n$,
we define new random variables $A_i \triangle A_j$
that are one when $A_i=A_j$ and zero otherwise.
Denote by $\braket{A}$
the expected value of a random variable $A$. 
The {\it full correlation vector x} based on $A_1,...,A_n$ is 
the vector of length $N=n + n(n-1)/2$ given by the expected values:
$$
x = ( \braket{A_i},
\braket{ A_i \triangle A_j})
\equiv
( \braket{A_i}_{1 \leq i \leq n} , 
\braket{ A_i \triangle A_j}_{1 \leq i < j \leq n}).
$$
Note that each element of the above vector lies between zero and one.
Now consider any vector $x=(x_1,...,x_n,x_{12},...,x_{n-1,n}) \in [0,1]^N$ 
indexed as above, which we will call an {\it outcome}.
We consider two related computational questions:

\vspace{3mm}
\noindent
{\bf Recognition.} {\it When is
an outcome $x$ a full correlation vector?}

\vspace{3mm}
\noindent
{\bf Optimization.} {\it For any $c \in R^N$ what is the maximum
value of $c^T x$ over all possible full correlation vectors $x$?}
\vspace{3mm}

It turns out that the recognition problem is NP-complete, and
the optimization problem is NP-hard. This follows from the 
fact that the set of full correlation vectors is in fact
the cut polytope $CUT_{n+1}$ defined on the complete graph $K_{n+1}$.
This polytope is defined as the convex hull of the $2^N$ full
correlation vectors obtained by deterministically setting
each random variable $A_i$ to either zero or one.
For details of the above and other facts about cut polytopes,
see the book by Deza and Laurent \cite{DL97a}.
For a vector $u=(u_1,...,u_d)$ the $L_1$-norm of $u$ is given
by $\| u \|_1 =\sum_{i=1}^d | u_i |$.
We have the following well-known characterization of the cut polytope.

\vspace{3mm}
\noindent
{\bf $L_1$-characterization of full correlation vectors.}
\par\noindent
{\it The following two statements are equivalent: 
\begin{itemize} 
\item 
An outcome $x \in [0,1]^N$ is a full correlation vector.
\item
There exist vectors $u^i \in R^{d}$,
$1 \le i \le n$,
$d \le N$,
for which
$$
x_i = \| u^i \|_1, ~~~~~
x_{ij} =  \| u^i - u^j \|_1 ,~~~1 \le i < j \le n.
$$
\end{itemize}
}
\par\noindent
Full correlation vectors provide an adequate model for correlations
obtained in physical experiments at the classical level.
Let us call the random variables observables.
For example, with $n=3$, $A_1, A_2, A_3$ could obtain the value one
if a given McGill student has blond hair, weighs more than 80 kg or
is more than 180cm high, respectively. We could obtain a 
full correlation vector by determining these three observables
for all McGill students.

In a quantum setting, things are very different.
Firstly, it is difficult to apply the above model directly
since at the quantum level it may be impossible to measure
directly different observables for a given particle.
Therefore the above model is replaced by a bipartite
setting where the 0/1 random variables (observables) are
labelled $A_1,...,A_m$ and $B_1,...,B_n$ respectively.
 
The {\it (bipartite) correlation vector x} based on 
random variables $A_1,...,A_m$
and $B_1,...,B_n$ is
the vector of length $M=m+n + mn$ given by the expected values:
\begin{eqnarray}
x &=& ( \braket{A_i}, \braket{B_j},
\braket{ A_i \triangle B_j}) \nonumber \\
&\equiv&
( \braket{A_i}_{1 \leq i \leq m} ,
 \braket{B_j}_{1 \leq j \leq n} ,
\braket{ A_i \triangle B_j}_{1 \leq i \leq m, 1 \leq j \leq n}).
\label{bcv}
\end{eqnarray}

As we will be concerned only with the bipartite case, we will
simply use the term correlation vector where no confusion arises.
As before, we call any vector $x \in [0,1]^M$ indexed as in
(\ref{bcv}) an {\it outcome}.
Again we may define a polytope by considering the convex hull
of the $2^{m+n}$ correlation vectors formed by letting each
of the $m+n$ random variables take value either zero or one.
This polytope is called the Bell polytope $B_{m,n}$ and
was apparently first considered by Froissart \cite{Fro-NC81}.
It turns out the membership and optimization problems given above
are still NP-complete and NP-hard respectively (for references, 
see, e.g., \cite{AII06a} ). The characterization theorem generalizes
in a natural way. 

\vspace{3mm}
\noindent
{\bf $L_1$-characterization of bipartite correlation vectors.}
\par\noindent
{\it The following two statements are equivalent:
\begin{itemize}
\item
An outcome $x \in [0,1]^M$ is a bipartite correlation vector.
\item
There exist vectors $u^i, v^j \in R^{d}$, 
$1 \le i \le m$, $1 \le j \le n$,
$d \le M$,
for which
$$
x_i = \| u^i \|_1, ~~~~~x_{m+j} = \| v^j \|_1, ~~~~~
x_{ij} =  \| u^i - v^j \|_1 .
$$
\end{itemize}
}
\par\noindent
The Bell polytope has been much studied.
Valid inequalities for the $B_{m,n}$ are often called {\it Bell inequalities}, although here we will reserve
this term for the facets of $B_{m,n}$.
These inequalities have been studied by many researchers, see for
example \cite{CG04a}, \cite{Pi91a}, \cite{WW01a}.  
The  {\it CHSH inequality}
is the only non-trivial facet of $B(2,2)$ and is given by
$$
\braket{ A_1 \triangle B_1} -\braket{ A_1 \triangle B_2} - \braket{ A_2 \triangle B_1} - \braket{ A_2 \triangle B_2}
\leq 0 
$$
or equivalently
\begin{equation}
x_{11} - x_{12} - x_{21} - x_{22} \leq 0.
\label{chsh}
\end{equation}
Although few Bell inequalities were known until recently,
much is known about facets of the cut polytope, including several
large classes of facets.
In \cite{AIIS05a} a method is given to generate Bell inequalities from
facets of the cut polytope, producing a large number of new
inequivalent Bell inequalities.

The correlation vector with $m=2, n=2$ given by
\begin{equation}
x = ( {1 \over 2}, {1 \over  2} , {1 \over 2} , {1 \over 2} , {{2 + \sqrt 2 } \over 4}  , 
{{2 - \sqrt 2 } \over 4} , {{2 - \sqrt 2 } \over 4} , {{2 - \sqrt 2 } \over 4} )
\label{viol}
\end{equation}
clearly violates the CHSH inequality (\ref{chsh}), 
so it follows there is no joint distribution function
for the four random variables. This correlation vector cannot arise as the result
of an experiment for which the rules of classical physics apply.
An outstanding prediction of quantum theory, apparently confirmed by numerous experiments,
is that this correlation vector can arise from observations at the quantum level.
This fact has lead to many surprising applications in quantum information theory, see
for example Cleve et al.\cite{CHTW04a}. 
It raises the issue of whether there is a good characterization
of such {\it quantum correlation vectors}, the topic of the rest of the paper.

\section{Quantum correlations}\label{section: quantum correlations}
\par\noindent
The postulates of quantum theory give a complete statistical description of the outcome
of experiments at the quantum level. A two party quantum
correlation experiment
can be described by a quantum state and
set of observables $A_1 ,...,A_m , B_1, ... B_n$ on a bipartite Hilbert space.
It is assumed the two parties are spatially separated and that the 
observations are performed essentially simultaneously,
so that there is insufficient time for the parties to communicate.
For a given experimental outcome,
the vector $x$ defined by (\ref{bcv}) is
called a {\it quantum correlation vector}.
The description given by the postulates does not appear to provide any tractable
method to answer the recognition, optimization and characterization questions
when applied to quantum correlation vectors. Such answers are provided, however, for one
important case by a theorem of Tsirelson. 
A {\it quantum correlation function } is a vector $y \in R^{mn}$
defined by taking the last $mn$ coordinates of a quantum correlation vector, i.e.,
\begin{equation}
y= (\braket{ A_i \triangle B_j})
\equiv
(\braket{ A_i \triangle B_j}_{1 \leq i \leq m, 1 \leq j \leq n}).
\label{qcv}
\end{equation}

\vspace{3mm}
\par\noindent
{\bf Tsirelson's Theorem (0/1 version)\cite{Ci80a} \cite{Ts87a}.}
\par\noindent
{\it
The following three statements 
are equivalent:
\begin{itemize}
\item
$y= (\braket{ A_i \triangle B_j}) \in [0,1]^{mn}$ is a quantum correlation function.
\item
$x = ( 1/2, 1/2, ... , 1/2,
\braket{ A_i \triangle B_j})\in [0,1]^M$ is a quantum correlation vector.
\item
There exist vectors $u^i, v^j \in R^{d}$, 
$1 \le i \le m$, $1 \le j \le n$,
$d \le m+n$,
for which
\begin{eqnarray*}
x_i = \| u^i \| = {1 \over 2},~~ x_{m+j} = \| v^j \| = {1 \over 2}, ~~  x_{ij} =   \| u^i - v^j \| .
\end{eqnarray*}
where $\| u \| \equiv u^T u $.
\end{itemize}
}
\vspace{3mm}
\par\noindent
We call an experimental outcome {\it unbiased } if for all $i$ and $j$
we have $\braket{ A_i} = \braket{ B_j } = 1/2$, otherwise it is {\it biased}. 
A remarkable result implied by this theorem is that 
the recognition and optimization problems for correlation functions
and unbiased 
quantum correlation vectors 
can be solved in polynomial time by
semidefinite programming(SDP).
Using the theorem, we can verify that (\ref{viol}) is
a quantum correlation vector by exhibiting the vectors:
\begin{eqnarray}
u^1 & = & ( {1 \over 2} , 0 , {1 \over 2} ) ,~~
u^2 = ( 0, {1 \over 2} , {1 \over 2} ) ,~~ \nonumber \\
 v^1 & = & ( {-1 \over {2 \sqrt 2 }},{ 1 \over {2 \sqrt 2 }}, {1 \over 2} ) ,~~
v^2 = ( {1 \over {2 \sqrt 2 }},{1 \over {2 \sqrt 2 }}, {1 \over 2} ).
\label{viol1}
\end{eqnarray}
Furthermore, it can be verified by SDP that this is the maximum
violation of (\ref{chsh}), although in this case Tsirelson \cite{Ci80a}
has provided an
analytic proof. 
These maximum quantum violations
have many interesting applications, see e.g. \cite{CHTW04a}.
The 
maximum quantum violation of any Bell inequality (like CHSH)
that does not have terms involving 
the expectations $\braket{A_i}$ or $\braket{B_j}$
can likewise be found by using SDP. 
Unfortunately, most of the Bell inequalities produced
recently \cite{AIIS05a} do not satisfy these conditions.
For these inequalities the maximum quantum violation may only be achieved
by a biased quantum correlation vector, and
the above method cannot be directly applied.

Tsirelson's theorem may not hold for
experimental outcomes that are biased. Consider the outcome for $m=n=1$
given by $x=(3/4, 3/4, 3/4 )$. If we set $u^1 = (  \sqrt 3 / 4, 3/4)$
and $v^1 =  (- \sqrt 3 / 4, 3/4 )$ then
$$
x_1 = \| u^i \|,~~ x_{2} = \| v^1 \|, ~~  x_{12} =   \| u^1 - v^1 \| ,
$$
and the corresponding vector $y=(3/4)$ is obviously a 
quantum correlation function.
However $x$ is not a quantum correlation vector
because it violates the no-signalling condition.
This condition derives from the fact that the expectations
$\braket{A_i}, 1 \le i \le m$ should be the same regardless of
which measurement $j$ the other party decides to make, due to
the spatial separation of the two parties. Similar conditions
should hold for the expectations
$\braket{B_j}$.
It is shown in \cite{AII06a} that a vector
$x$ satisfies the {\it no-signalling condition} if and only if
it belongs to the {\it rooted semimetric polytope} defined by
the inequalities:
\begin{eqnarray}
x_i + x_j + x_{ij} \le 2,~~~
x_i + x_j - x_{ij} \ge 0,~~~ \nonumber \\
x_i - x_j + x_{ij} \ge 0,~~~
-x_i + x_j + x_{ij} \ge 0.
\label{rsm}
\end{eqnarray}
It is easy to see that unbiased quantum correlation vectors 
satisfy the
no-signalling condition. 
However, the vector $x=(3/4, 3/4, 3/4 )$ violates the first of these inequalities.

It is tempting to conjecture that an outcome $x$
is a quantum correlation vector if it satisfies the no-signalling
conditions 
(\ref{rsm}) and the corresponding vector $y$ is a quantum correlation function.
However, consider the vectors
$$
x = ( {1 \over 2 \sqrt 2}, {1 \over  2 \sqrt 2} , 
{1 \over 2} , {1 \over 2} , {{2 + \sqrt 2 } \over 4}  ,
{{2 - \sqrt 2 } \over 4} , {{2 - \sqrt 2 } \over 4} , {{2 - \sqrt 2 } \over 4} )
$$
$$
y = ({{2 + \sqrt 2 } \over 4}  ,
{{2 - \sqrt 2 } \over 4} , {{2 - \sqrt 2 } \over 4} , {{2 - \sqrt 2 } \over 4} ).
$$
The outcome $x$ satisfies (\ref{rsm}), and $y$ is a quantum correlation function,
as shown by the vectors given in (\ref{viol1}). Nevertheless, it is proved
in \cite{AI07a} that $x$ is {\em not} a quantum correlation vector.
Perhaps even more surprising is a non-quantum outcome they exhibit 
for the case $m=n=3$:
\begin{eqnarray*}
x_i={1 \over 3},~~1 \le i \le 6,~~ x_{11}=x_{22}=0,~~
x_{ij}=
 {2 \over 3}~~  \\ for~all~other~1 \le i < j \le 3.
\end{eqnarray*}
This gives an outcome $x$ which satisfies (\ref{rsm}) and for
which the corresponding correlation function can even be
obtained classically. For example with vectors
\begin{eqnarray*}
&\ &u^1 = v^1 = (0,0,0,1/3),~~ u^2 = v^2 = (0,0,1/3,0),~~ \\
& \ &u^3 = (1/3 , 0,0,0),~~ v^3 = (0, 1/3, 0, 0 )
\end{eqnarray*}
we have
$$
 x_{ij} =  \| u^i - v^j \|_1~~~~1 \le i < j \le 3.
$$
and can use the $L_1$ characterization theorem given in the previous section.

Even though Tsirelson's theorem does not give a characterization 
of quantum correlation vectors, it can be extended to give a necessary condition
that can be combined with the no-signalling condition.

\vspace{3mm}
\par\noindent
{\bf Necessary conditions for quantum correlation vectors \cite{AII06a}.}
\par\noindent
If {\it $x = ( \braket{A_i}, \braket{B_j},
\braket{ A_i \triangle B_j}) \in [0,1]^M$ is a quantum correlation vector
then
\begin{itemize}
\item
$x$ must satisfy the no-signalling conditions (\ref{rsm}), and 
\item
There exist vectors $u^i, v^j \in R^{d}$,
$1 \le i \le m$, $1 \le j \le n$,
$d \le m+n$,
for which
\[
x_i = \| u^i \|, ~~ x_{m+j} = \| v^j \|, ~~  x_{ij} =   \| u^i - v^j \| .
\]
where $\| u \| \equiv u^T u $.
\end{itemize}
}
\vspace{3mm}
\par\noindent
Using this theorem, it can be shown that in the two previous examples
the outcomes are not quantum correlation vectors.
It also provides an efficient means of bounding the maximum quantum violation
of general Bell inequalities by semidefinite programming (SDP).
The strength of the no-signalling conditions
with respect to various known Bell inequalities is discussed further in
\ref{section: maximum violation}.

It has recently been shown that the above conditions are not sufficient:
in Doherty et al. 
\cite{DLTW} and Navascu\'es et al. \cite{NPA} a hierarchy of SDPs is
presented 
which in some cases gives much
tighter bounds.
A characterization of quantum correlation vectors
is given by an infinite hierarchy of conditions, each of which can be
tested by solving an SDP. This work is reviewed in Section \ref{MV}.
It is still, however, an open problem to determine whether or
not there is a polynomial time algorithm to determine
whether a given vector is a
quantum correlation vector.

\section{A proof of Tsirelson's theorem}\label{section: Tsirelson's Theorem}
\par\noindent
In this section we give an elementary description of 
the easy direction in Tsirelson's proof.
For the proof, it is convenient to
let the observables $A_1 , ... , A_m , B_1 , ... , B_n$
take values $\pm 1$ rather than 0/1, and to consider the products
$\braket{A_iB_j}$ rather than the differences
$\braket{A_i \triangle B_j}$. An outcome is now given by
\begin{eqnarray*}
x & = & ( \braket{A_i}, \braket{B_j},
\braket{ A_i B_j}) \\
&\equiv &
( \braket{A_i}_{1 \leq i \leq m} ,
 \braket{B_j}_{1 \leq j \leq n} ,
\braket{ A_i B_j}_{1 \leq i \leq m, 1 \leq j \leq n}),
\end{eqnarray*}
and is called a quantum correlation vector if it
can result from a quantum experiment. Similarly
we redefine a quantum correlation function.
In this section we 
use the ket-bra notation  where $\ket{u}$ denotes a
(possibly complex) vector, $\bra{u}$ 
denotes the transpose of its complex conjugate,
and $\braket { u | v}$ denotes inner product.
Using these new notations, the theorem takes the
following equivalent form  \cite{Ci80a,Ts87a}.

\begin{Theorem}[\bf Tsirelson's Theorem ($\pm 1$ version)]\label{theorem: Tsirelson}
For real $m \times n$ matrix $c_{ij}$, the followings are equivalent:
\begin{itemize}
\item[(1)]
$y= (c_{ij}) \in [-1,1]^{mn}$ is a quantum correlation function.
\item[(2)]
$x = ( 0, 0, ... , 0,
c_{ij})\in [-1,1]^{m+n+mn}$ is a quantum correlation vector.
\item[(3)]
There exist two sets of unit vectors $\{ u^i \}_{i=1}^m, \{v^j \}_{j=1}^n \in \mathbb{R}^{d}$,
$d \le m+n$,
for which
\[
c_{ij} =  \braket{u^i | v^j}.
\]
\item[(4)]
There exist two sets of vectors $\{ u ^i \} _{i=1}^m$ and $\{ v^j \}_{j=1}^n$ on $\mathbb{R}^{\min (m, n)}$ such that
they satisfy $|u^i| \le 1$, $|v^j| \le 1$ for all $i$ and $j$,
\begin{equation}
c_{ij} = \braket{ u^i | v^j}. 
\end{equation} 
\item[(5)]
There exist a state on a composite Hilbert space $\Hi _A \otimes \Hi _B$ and 
two sets of Hermitian operator $\{ A_i \}_{i=1}^m$ on a Hilbert space $\Hi _A$ and
$\{ B_j \}_{j=1}^n$ on a Hilbert space $\Hi _B$  such that $|A_i| \le I$, $|B_j| \le I$, and
\begin{equation}\label{eq: Tsirelson's theorem 4}
c_{ij} = \Tr (A_i \otimes B_j \rho). 
\end{equation}
\end{itemize}
\end{Theorem}

Proof: If (1) holds, some quantum correlation vector $x'$
must be consistent with $y$. By switching the outcomes +1 and -1 we
see that the vector $x''$ formed by setting $x_i'' = -x_i',~ i=1,...,m+n$
and otherwise setting $x_{ij}'' = x_{ij}'$ is also a quantum correlation vector.
The implication $(1) \Rightarrow (2)$ then follows from the convexity of the set
of quantum correlation vectors by setting $x= (x' + x'')/2$. 

The implication $(2) \Rightarrow (3)$ follows from the postulates of quantum
theory. Indeed, corresponding to the given
quantum correlation vector there must exist observables
$A_1,...,A_m$ on a Hilbert space
$\Hi _A$, observables $B_1,...,B_n$ on a Hilbert space
$\Hi _B$, and a pure quantum state $\ket{\psi}$ given as
a unit vector on 
$\Hi _A \otimes \Hi _B$, where $\otimes$ denotes tensor (Kronecker) product.
For $1 \le i \le m$ and $1 \le j \le n$,
let $\ket {a^i}=   A_i \otimes I_B    \ket {\psi }$
and $\ket {b^j}=   I_A \otimes B_j  \ket {\psi }$.
Then  $\ket {a^i}$ and $\ket {b^j}$ are (possibly complex) 
unit vectors of length, say, $t$, such that $\braket{a^i | b^j}=\braket{A_iB_j}$.
We may replace them with real vectors $u^i$ and 
$v^j$ of length $2t$ by writing the
real and complex coefficients as separate coordinates, maintaining the same
values of the inner products. 
The set of $m+n$ real vectors $u^1,...,u^m,v^1,...,v^n$
have all the properties of part (2)
except possibly their dimension $2t > m+n$. However the unit
vectors lie in a subspace of dimension $d \le m+n$, which preserves their inner 
products. We can also prove $(5) \Rightarrow (4)$ in the same way.

Now, we prove $(4) \Rightarrow (5)$ and $(3) \Rightarrow (1)$.
Suppose the condition $(4)$ (or the condition $(3)$) holds, and $\xi=\min \left ( \dim {\rm span} \{ u^i \}_i^m,  \dim {\rm span} \{ v^j \}_j^n \right )$ 
(or $\xi=d$ for the condition $(3)$).  
We choose a Hilbert space $\Hi$ whose dimension is $2^\nu $, 
where $\nu= \frac{\xi}{2}$
for even $\xi$, 
and $\nu= \frac{\xi-1}{2}$
for odd $\xi$. 
Then, there exists an irreducible representation of the Clifford algebra on $\Hi$,
and we can choose a set of $\xi$ 
Hermitian operators $\{ X_k \}_{k=1}^{\xi} $ as an irreducible 
representation of a generator set of the Clifford algebra (see  \ref{section Clifford});
that is, $\{ X_k \}_{k=1}^{\xi} $ satisfies the following anti-commutation relation:
\begin{equation}\label{anti-commutation}
X_iX_j+X_jX_i= 2\delta _{ij}I.
\end{equation}
Then, from the above relation,
Hermitian operators $\{ X_k \otimes X_k \}_{k=1}^{\xi} $ commute 
with each other. Thus, they have simultaneous eigenvectors,
and moreover, their corresponding eigenvalues are  $\pm 1$. 
Therefore, there exists a state $\ket{\Psi} \in \Hi _A \otimes \Hi _B$ 
which is a simultaneous eigenvector of $\{ X_k \otimes X_k \}_{k=1}^{\xi}$ satisfying 
\begin{equation}\label{eq: extended eigen eq}
X_k \otimes X_k \ket{\Psi} = (-1)^{a_k} \ket{\Psi}
\end{equation} 
where $a_k$ is $0$ or $1$. 
Then, $\ket{\Psi}$ and $\{ X_k \}_{k=1}^{\xi}$ satisfies
\begin{eqnarray}\label{eq: orthogonality}
& \quad & \bra{\Psi} X_k \otimes X_l \ket{\Psi} \nonumber \\ &=& 
\frac{1}{2}  \{ (-1)^{a_l} \bra{\Psi} X_kX_l \otimes I \ket{\Psi} \nonumber \\
&\quad &  \quad + (-1)^{a_l} \bra{\Psi} X_lX_k \otimes I \ket{\Psi} \} \nonumber\\
&=& \frac{1}{2}(-1)^{a_l} \bra{\Psi} (X_kX_l+X_lX_k) \otimes I\ket{\Psi} \nonumber \\
&=& (-1)^{a_l}\delta _{kl} I, 
\end{eqnarray}
where we use Eq.(\ref{eq: extended eigen eq}) in the first equality and Eq.(\ref{anti-commutation}) in the third equality.
Finally, we define $A_i$ and $B_j$ as 
\begin{eqnarray}\label{definition of A_k and B_l}
&& A_i = \sum _{k=1}^{\xi} (u'^i)_k X_k \label{def: A} \nonumber \\
&& B_j = \sum _{k=1}^{\xi} (-1)^{a_k} (v'^j)_k X_k \label{def: AB'},
\end{eqnarray}
where $u'^i$ and $v'^j$ are vectors derived 
by projecting $u^i$ and $v^j$ onto ${\rm span} \{ u^i \}_{i=1}^m$ in the case of $\dim {\rm span} \{ u^i \}_{i=1}^m \le \dim {\rm span} \{ v^j \}_{j=1}^n$ 
and onto ${\rm span} \{ v^j \}_{j=1}^m$ otherwise (for condition $(3)$, we can just choose $u'^i=u^i$ and $v'^i=v^i$). 
By their definition, we can easily see that $A_i^2=|u'^i|^2I$ and $B_j^2=|v'^j|^2I$.
This fact guarantees that $|A_i|\le I$ and $|B_j|\le I$ (for the condition $(3)$, $A_i$ and $B_j$ take eigenvalues $\pm 1$).
Finally, we can calculate $\bra{\Psi} A_k \otimes B_l \ket{\Psi}$ as follows: 
\begin{eqnarray}\label{e1}
&\quad & \bra{\Psi} A_i \otimes B_j \ket{\Psi} \nonumber \\
&=& \sum _{kl} (-1)^{a_l}(u'^i)_k (v'^j)_l \bra{\Psi}X_k \otimes X_l \ket{\Psi} \nonumber \\
&=& \sum _{kl} (-1)^{2a_l}(u'^i)_k (v'^j)_l \delta _{kl} \nonumber \\
&=& \left <u'^i | v'^j \right > \nonumber \\
&=& \left <u^i | v^j \right >
\end{eqnarray}
where we use Eq.(\ref{eq: orthogonality}) in the second equality. \hfill $\square$

In the above proof of $(4) \Rightarrow (5)$ and $(3) \Rightarrow (1)$, 
in order to satisfy the equation $\bra{\Psi} A_i \otimes B_j \ket{\Psi} = \left <u^i | v^j \right >$,  
Tsirelson used the simultaneous eigenvector $\ket{\Psi}$ corresponding to
eigenvalue ``1'' of $\{ X_k \}_{k=1}^{\xi}$ \cite{Tsi-HJS93}.
With $A_i$ and $B_j$ defined by Eq.(\ref{definition of A_k and B_l}) and $a^k=0$,
we showed that we can choose an arbitrary simultaneous eigenvector of $\{ X_k \otimes X_k \}_{k=1}^{\xi}$ 
by slightly changing the definition of $A_i$ and $B_j$ from Tsirelson's original definition. 
As we will see later, this modification of the proof is actually important for finding an explicit optimal construction 
of $A_i$ and $B_j$.

\section{Representing two dimensional vectors}
\par\noindent
In order to illustrate the concepts, we give a construction
of a quantum setting that can realize the inner products of two dimensional vectors. 
Let
$u^1 , ... , u^m$
and $v^1 , ... , v^n$ be $m+n$ unit vectors in $R^2$
and let $z=( z_1 , z_2 )$. Define
$$ 
C(z) = \left[ \matrix {
               \matrix { z_1 \cr z_2 }
               \matrix { z_2 \cr -z_1 }
                     } \right] .
$$
Now define observables
$$
A_i = C( u^i ),\ \ \ \ B_j = C ( v^j ) ,\ \ \ \  i=1,...,m\ \ j=1,...,n.
$$
Let $\ket{\Psi} = [1/ { \sqrt 2 } ,\  0,\  0,\  1/ { \sqrt 2 } ]^T $ which corresponds to the
state $ { \ket{00}  +  \ket{11}   \over { \sqrt 2 } }.$
Then it is easy to verify that Eq.(\ref{e1}) holds. For example, with
\begin{eqnarray*}
u^1 &=& [ 1, 0 ]^T ,\ \  u^2 = [0, 1]^T ,\\  
v^1 &=& [1/ { \sqrt 2 } ,\ 1/ { \sqrt 2 } ]^T ,\ \ 
v^2 = [1/ { \sqrt 2 } ,\ -1/ { \sqrt 2 } ]^T
\end{eqnarray*}
we may verify, for instance, that
$$
\braket{ u^2 | v^1}  \ =\   \braket { \Psi  |  C ( u^2 ) \otimes C ( v^1 )  |  \Psi } 
\ =\  {1 \over { \sqrt 2 } }.
$$
Incidentally, this gives a quantum setting for the maximum violation of the CHSH inequality
described in Section \ref{class}, when the inequality 
is restated in $\pm 1$ terms.

Here is how the quantum setting is obtained. We start with
$$
X_1 =  \left[ \matrix {
               \matrix { 1 \cr 0 }
               \matrix { 0 \cr -1 }
                     } \right] \ \ \ 
 X_2 =  \left[ \matrix {
               \matrix { 0 \cr 1 }
               \matrix { 1 \cr 0 }
                     } \right] .
$$
It is easy to verify that the anti-commutation relation (\ref{anti-commutation}) 
holds (that is, $X_1, X_2$ are a representation of the generators of the Clifford algebra $Cl(\mathbb{R}^2)$) and that $C(z) = X(z)$ as given
by (\ref{definition of A_k and B_l}) with $a_k=0$. To get the state $\ket{\Psi}$ we first construct the operator
$A$ given by
$$
2 A \ =\   X_1 \otimes X_1 \ +\    X_2 \otimes X_2
\ =\  \left[ \matrix {
      \matrix { 1 \cr  0 \cr  0 \cr  1 }
      \matrix { 0 \cr -1 \cr  1 \cr  0 }
      \matrix { 0 \cr  1 \cr -1 \cr  0 }
      \matrix { 1 \cr  0 \cr  0 \cr  1 }
         } \right] . 
$$
Using Maple we find that $A$ has maximum eigenvalue $1$ with corresponding
eigenvector $ [1 ,\  0,\  0,\  1]^T $. When normalized, this eigenvector gives
the state $\ket{\Psi}$ above.

\section{Construction of a state and observables which represent correlation functions}\label{problem setting}
In this section, we give a general construction,
that is, states and operators which can represent all 
correlation functions.
By using the theory of stabilizer states, especially, the {\it binary representation} of stabilizer formalism, we show that we can always choose a 
standard singlet for a state to represent all correlation functions.   
We believe this is the first time an explicit construction has been given.
For the convenience of readers, we give a review of the
binary representation of stabilizer formalism in  \ref{section: binary representation}.

We are interested in the construction of the states $\ket{\Psi}$ and observables $A_i$ and $B_j$ 
satisfying $|A_i| \le I$, $|B_j| \le I$ (or $|A_i| = |B_j| = I$), and
\begin{equation}\label{eq: main condition}
\bra{\Psi}A_i\otimes B_j\ket{\Psi}=\braket{u^i | v^j}
\end{equation}
in the case when two sets of vectors $\{ u^i \}_{i=1}^m$ and $\{ v^j \}_{j=1}^n$ satisfying 
the fourth condition (or the third condition) of Theorem \ref{theorem: Tsirelson} are given.
As we have seen in the proof of the theorem, 
suppose $\xi=\min \left ( \dim {\rm span} \{ u^i \}_i^m,  \dim {\rm span} \{ v^j \}_j^n \right )$ for 
the fourth condition, 
and $\xi=d$ for the third condition. 
Then, these vectors are represented in 
a Hilbert space $\Hi$ whose dimension is $2^\nu $ ($\nu= \frac{\xi}{2}$
for even $\xi$, 
and $\nu= \frac{\xi-1}{2}$
for odd $\xi$) as Eq.(\ref{definition of A_k and B_l})
by using a representation of the generators of the Clifford algebra $\{ X_k \}_{k=1}^{\xi}$ and 
their simultaneous eigenvector $\ket{\Psi}$.
Therefore, in order to construct $\ket{\Psi}$, $A_i$ and $B_j$, 
we need to know how we can find a simultaneous eigenvector 
$\ket{\Psi}$ corresponding to $\{ X_k \}_{k=1}^{\xi}$.
In what follows, we give an explicit way to do it.


First, we consider the case when $\xi$ is even; that is, $\xi=2\nu$; 
In this case, we can choose the Weyl-Brauer matrices as an irreducible representation of the
generators of the Clifford algebra on $\Hi _A$ 
and $\Hi _B$ as follows (see \ref{section Clifford}):
\begin{eqnarray}
X_k &\stackrel{\rm def}{=}& \overbrace{Z\otimes \cdots \otimes Z}^{k-1}\otimes X \otimes \overbrace{I\otimes \cdots \otimes I}^{\nu-k}, \nonumber \\
X_{k+\nu} &\stackrel{\rm def}{=}& -\overbrace{Z\otimes \cdots \otimes Z}^{k-1}\otimes Y \otimes \overbrace{I\otimes \cdots \otimes I}^{\nu-k}, \label{definition X_i}
\end{eqnarray} 
where $X$, $Y$ and $Z$ are the Pauli matrices, and an index $k$ satisfies $1 \le k \le \nu$ \cite{Weyl}. 
Since $\{ X_k \}_{k=1}^{2\nu}$ is a set of Hermitian operators which are composed of tensor products of Pauli operators satisfying
Eq.(\ref{anti-commutation}), 
$\{ X_k \otimes X_k \} _{k=1}^{2\nu}$ generate a commutable subgroup of 
the Pauli Group \cite{G97,NDM04,HDERNB06} on $2\nu$ qubits $\mathcal{P}_{2\nu}$; we write this commutable subgroup as $S_{2\nu}$.  
Note that Eq.(\ref{definition X_i}) guarantees independence of the generator 
$\{ X_k \otimes X_k \}_{i=1}^{2\nu}$.
Since $\ket{\Psi}$ is defined as the simultaneous eigenvector of $\{ X_k \otimes X_k \}_{k=1}^{2\nu}$ 
corresponding to an eigenvalue $1$ for all $X_k \otimes X_k$,
$\ket{\Psi}$ is nothing but a stabilizer state corresponding to the independent and commutative Pauli group elements $\{ X_k \otimes X_k \}$. 
Here, in a $n$-qubit Hilbert space, a stabilizer state  $\ket{\Psi}$ of commutative and independent Pauli group elements $\{ M_k \}_{k=1}^n$  
is defined as a state which satisfies $M_i\ket{\Psi} = \ket{\Psi}$ for all $i$ \cite{NDM04,HDERNB06}.
In this case, a set of $2^n$ commuting Pauli operators 
$S \stackrel{\rm def}{=} \{ M \in  \mathcal{P} _n | M\ket{\Psi} = \ket{\Psi} \}$ is called a stabilizer group
of the state $\ket{\Psi}$; thus, $\{ M_k \}_{k=1}^n$ are independent generators of the commutative group $S$.  
Conversely, a set of $2^n$ commuting Pauli operators all of which have real overall phase $\pm 1$ uniquely 
define a corresponding stabilizer state by the equations $M\ket{\Psi} = \ket{\Psi}$ for all $M \in S$.      
Thus, our task is to find a concrete expression for a stabilizer state $\ket{\Psi}$ corresponding to 
the stabilizer group $S_{2\nu}$ which is generated by $\{ X_k \otimes X_k \}_{k-1}^{\xi}$. 

For the case when $\xi=2\nu +1$,
we can choose an irreducible representation of the
generators of the Clifford algebra on $\Hi _A \otimes \Hi _B$
as $\{ X_k \}_{k=1}^{2\nu +1}$ where $\{ X_k \}_{k=1}^{2\nu}$ are defined by Eq.(\ref{definition X_i}) and 
$X_{2\nu +1} \stackrel{\rm def}{=} \overbrace{Z \otimes \cdots \otimes Z}^{\nu}$ (see \ref{section Clifford}).
Since the equality $X_{2\nu +1} = (-i)^{\nu} \Pi _{k=1}^{2\nu} X_k$ holds,
we derive the equality $X_{2\nu +1} \otimes X_{2\nu +1} = (-1)^{\nu}\Pi _{k=1}^{2\nu} X_k \otimes  X_k \in S_{2\nu}$.
Thus, a stabilizer state  $\ket{\Psi}$ of the stabilizer group $S_{2\nu}$ 
automatically satisfies $X_{2\nu +1}\otimes X_{2\nu +1}\ket{\Psi} = (-1)^{\nu+\sum _{k=1}^{2\nu} a_k}\ket{\Psi}$.
Therefore, by defining $a_{2\nu +1}=\nu+\sum _{k=1}^{2\nu} a_k$, 
the state $\ket{\Psi}$ is nothing but the state which satisfies $\bra{\Psi} A_i \otimes B_j \ket{\Psi} = \left <u^i | v^j \right >$ for $A_i$ and $B_j$ 
defined by  Eq.(\ref{definition of A_k and B_l}).

So far, we showed that the state $\ket{\Psi}$ satisfying Eq.(\ref{eq: main condition})
is a stabilizer state of the group $S_{2\nu}$ which is generated by $\{ X_k \otimes X_k \}_{k=1}^{2\nu}$.
In order to construct a simultaneous eigenstate of $\{ X_k \otimes X_k \}_{k=1}^{2\nu}$ defined by Eq.(\ref{definition X_i}), 
we find a graph state which is equivalent to $\ket{\Psi}$ under local Clifford operations
by using the binary representation of the stabilizer formalism,  
which was mainly developed in  \cite{G97} and \cite{NDM04} (See  \ref{section: binary representation}). 
Here, a graph state is a state in the standard form of stabilizer states \cite{HDERNB06}, and 
defined by means of an $n$-vertex graph $G$ with adjacency matrix $\{ \theta _{ij} \}_{ij}$ as the stabilizer state 
generated by commuting Pauli's operators $\{K_j \}_{j=1}^n $ defined as 
\begin{equation}
K_j = X^{(j)}\Pi_{k=1}^n \left ( Z^{(k)} \right )^{\theta _{kj}}. 
\end{equation}
Here $X^{(i)}$ and $Z^{(i)}$ are the Pauli operators having $X$ and $Z$ in the $i$th position of the tensor
product and the identity elsewhere, respectively. 
We note that the binary generator matrix $K$ corresponding to $\{ K_j \}_{j=1}^n$ can be written as the following simple form:
\begin{equation}\label{matrix graph state}
K= \stackrel{\rm def}{=} \left [
\begin{array}{c}
\theta \\
I
\end{array}
\right ]. 
\end{equation}

In the binary representation, the set of Pauli operators $\{ X_k \otimes X_k \}_{k=1}^{2\nu}$ defined in Eq.(\ref{definition X_i}) is written as 
\begin{eqnarray*}
X_k \otimes X_k &\leftrightarrow & 
(\overbrace{1 \cdots 1}^{k-1}\overbrace{0 \cdots 0}^{\nu-k+1}|\overbrace{1 \cdots 1}^{k-1}\overbrace{0\cdots 0}^{\nu-k+1}| \\
&\quad & \quad 
\overbrace{0 \cdots 0}^{k-1}1\overbrace{0\cdots 0}^{\nu-k}|\overbrace{0 \cdots 0}^{k-1}1\overbrace{0\cdots 0}^{\nu-k} )^T \\ 
X_{k+\nu} \otimes X_{k+\nu} &\leftrightarrow& (\overbrace{1 \cdots 1}^{k}\overbrace{0 \cdots 0}^{\nu-k}
|\overbrace{1 \cdots 1}^{k}\overbrace{0\cdots 0}^{\nu-k}| \\
&\quad & \quad  \overbrace{0 \cdots 0}^{k-1}1\overbrace{0\cdots 0}^{\nu-k}|\overbrace{0 \cdots 0}^{k-1}1\overbrace{0\cdots 0}^{\nu-k} )^T,
\end{eqnarray*}
where $1 \le k \le \nu$.
The generator matrix $E$ corresponding to a set of Pauli operators $\{ X_k \otimes X_k \}_{k=1}^{2\nu}$ can be written as
\begin{equation}
E \stackrel{\rm def}{=} \left ( 
\begin{array}{ccc}
T_{\nu-1} & T_{\nu} \\
T_{\nu-1} & T_{\nu} \\
I       & I \\
I & I
\end{array}
\right ),
\end{equation} 
where $T_{\nu -1}$ and $T_{\nu}$ are $\nu \times \nu$ matrices defined as
\begin{eqnarray*}
T_{\nu-1} & \stackrel{\rm def}{=} & \left ( 
\begin{array}{cccrc}
0 & 1 & 1 & \cdots & 1 \\
0 & 0  & 1 & \cdots & 1 \\
\vdots & \ddots  &\ddots & \ddots & \vdots \\
0 & 0 &\ & 0 & 1 \\
0 & 0 &\ &\cdots & 0
\end{array}
\right ), \\
T_{\nu} & \stackrel{\rm def}{=} & \left ( 
\begin{array}{clrc}
 1 & 1 & \cdots & 1 \\
 0  & 1 & \cdots & 1 \\
 \ddots  &\ddots & \ddots & \vdots \\
 0 &\ & 0 & 1 
\end{array}
\right ),
\end{eqnarray*}
and $I$ is a $\nu \times \nu$ identity matrix.

In order to find a graph state which is equivalent to $\ket{\Psi}$ under local Clifford operations,
we try to transform the above generator matrix $E$ into the form of Eq.(\ref{matrix graph state}),
which corresponds to a generator matrix of graph states, by changing the basis of the stabilizer and 
using local Clifford operations. 
As we can see in \ref{section: binary representation}, a change of the basis of a stabilizer corresponds to applying 
an invertible matrix from the right-hand side of $E$, 
and an application of local Clifford operations to the state $\ket{\Psi}$ 
is equivalent to applying a symplectic matrix from the left-hand side of $E$.
Following a discussion which is similar to that used in \cite{NDM04},
we can find and easily check the following equation:
\begin{equation} \label{eq: generator transform}
L \cdot  E \cdot  R_1 \cdot R_2 = F,
\end{equation} 
where $L$, $R_1$, $R_2$, $F$ are defined as
\begin{eqnarray}
&& L \stackrel{\rm def}{=} \left [
\begin{array}{cccc}
I & 0 & 0 & 0 \\
0 & 0 & 0 & I \\
0 & 0 & I & 0 \\
0 & I & 0 & 0
\end{array}
\right ], 
\quad 
R_1 \stackrel{\rm def}{=} \left [
\begin{array}{cc}
T_{\nu -1} & T_{\nu} \\
T_{\nu -1} & T_{\nu} \\
I & I \\
I & I
\end{array}
\right ], \nonumber \\
&& R_2 \stackrel{\rm def}{=} \left [
\begin{array}{cc}
I & T_{\nu} \\
T_{\nu} & I 
\end{array}
\right ], 
\quad 
F \stackrel{\rm def}{=} \left [
\begin{array}{cc}
0 & I \\
I & 0 \\
I & 0 \\
0 & I 
\end{array}
\right ].
\end{eqnarray}
We also easily see that $L$ is symplectic, $R_1$ and $R_2$ are invertible, 
and $F$ is in the form of Eq.(\ref{matrix graph state}); 
that is, $F$ is a generator matrix corresponding to a graph state. 
Actually, in comparison with Eq.(\ref{matrix graph state}), 
in this case, the adjacency matrix of the graph state corresponding to
$F$ is $\theta = \left [
\begin{array}{cc}
0 & I \\
I & 0 \\
\end{array}
\right ]
$. 
A way of constructing an arbitrary graph state is already known \cite{HDERNB06}.
Thus, we can easily see the graph state $\ket{\Psi'}$ corresponding to $F$ is 
\begin{equation}
\ket{\Psi'} = \bigotimes _{k=1}^{\nu} \frac{1}{\sqrt{2}}\left ( \ket{+\ 0}_{A_kB_k} \otimes \ket{-\ 1}_{A_kB_k} \right ),
\end{equation}
where $\ket{+} =\frac{1}{\sqrt{2}} \left ( \ket{0}+\ket{1} \right )$ 
and $\ket{-} =\frac{1}{\sqrt{2}} \left ( \ket{0} - \ket{1} \right )$
and that $\ket{\Psi'}$ is equivalent to $n$ singlets. 

As we can see in \ref{section: binary representation}, 
in Eq.(\ref{eq: generator transform}), 
applying $R_1$ and $R_2$ corresponds to changing the generators of 
a stabilizer group of $\ket{\Psi}$, 
and applying $L$ actually corresponds to applying a Clifford operation on 
$\ket{\Psi}$. As we can easily check, 
a binary symplectic matrix $L$ corresponds to a ``local'' Clifford operation  
$\mathcal{L} \stackrel{\rm def}{=} I \otimes \cdots \otimes I \otimes H \otimes \cdots \otimes H$,
where the first $n$-qubits belong to $\Hi_A$, the next $n$-qubits belong to $\Hi _B$,
and $H$ is a Hardmer operator.
Thus, the stabilizer groups corresponding to $\ket{\Psi}$ and $\mathcal{L}^{\dagger}\ket{\Psi'}=\mathcal{L}\ket{\Psi'}$
have the same binary representation $E = L^{-1}\cdot F \cdot R_2^{-1} \cdot R_1^{-1}$.
This fact does not guarantee $\ket{\Psi} =\mathcal{L}\ket{\Psi'}$. However, this fact guarantees that 
$\mathcal{L}\ket{\Psi'}$ is a simultaneous eigenvector of $\{ X_k \otimes X_k \}_{k=1}^{2\nu}$, which is a stabilizer of $\ket{\Psi}$. 
In fact, we can easily check that $\mathcal{L}\ket{\Psi'} = \bigotimes _{i=1}^{\nu} \frac{1}{\sqrt{2}}(\ket{++} + \ket{--})$ satisfies the following equations:
For $1 \le k \le \nu$, 
\begin{eqnarray*}
X_k \otimes X_k \mathcal{L}\ket{\Psi'} & = & \mathcal{L}\ket{\Psi'} \\
X_{k+\nu} \otimes X_{k+\nu} \mathcal{L}\ket{\Psi'} & = & - \mathcal{L}\ket{\Psi'} \\
X_{2\nu+1} \otimes X_{2\nu+1} \mathcal{L}\ket{\Psi'} & = &  \mathcal{L}\ket{\Psi'}, 
\end{eqnarray*}
where we only use $X_{2\nu+1}$ in the case  $\xi$ is odd.
Therefore, as we have already mentioned, we can set $a_k=0$ for $1 \le k \le \nu$ and $k=2\nu+1$, 
and $a_k =1$ for $\nu + 1 \le k \le 2\nu$, and we also define $A_i$ and $B_j$ by Eq.(\ref{definition of A_k and B_l}). 
Then, $\ket{\Psi''} \stackrel{\rm def}{=} \mathcal{L}\ket{\Psi'}$ satisfies
\begin{equation}
\bra{\Psi''} A_i \otimes B_j \ket{\Psi''} = \left < u^i | v^j \right >.
\end{equation}
Finally, as a result, by applying $\overbrace{H \otimes \cdots \otimes H}^{2\nu}$ to both $\ket{\Psi''}$ and $\{ X_k \otimes X_k \}_{k=1}^{2\nu +1}$,
we derive the following Theorem.  
\begin{Theorem}\label{thm2}
Given vectors $\{ u^i \} _{i=1}^m$ and $\{ v^j \}_{j=1}^n$ on $\mathbb{R}^{\min (m, n)}$ satisfying
$|u^i| \le 1$, $|v^j| \le 1$  for all $i$ and $j$, 
there are operators $\{ A_i \}_{i=1}^m$ and $\{ B_j \}_{j=1}^n$
on $2\nu$ qubits which satisfy $|A_i| \le I$, $|B_j| \le I$, and 
\begin{equation}
\bra{\Phi_+}^{\otimes \nu} A_i \otimes B_j \ket{\Phi_+}^{\otimes \nu} = \left < u^i | v^j \right >,
\end{equation} 
where 
$\ket{\Phi_+} \stackrel{\rm def}{=} \ket{00} + \ket{11}$, 
$\nu = \lfloor \frac{\xi}{2} \rfloor$, $\xi=\min \left ( \dim {\rm span} \{ u^i \}_i^m,  \dim {\rm span} \{ v^j \}_j^n \right )$, and
$A_i$ and $B_j $ are defined as 
\begin{eqnarray*}
&& A_i = \sum _{k=1}^{\xi} (u'^i)_k X_k \nonumber \\
&& B_j = \sum _{k=1}^{\xi} (-1)^{a_k} (v'^j)_k X_k.
\end{eqnarray*}
In the above,
$u'^i$ and $v'^j$ are vectors derived 
by projecting $u^i$ and $v^j$ onto ${\rm span} \{ u^i \}_{i=1}^m$ in the case of $\dim {\rm span} \{ u^i \}_{i=1}^m \le \dim {\rm span} \{ v^j \}_{j=1}^n$ 
and onto ${\rm span} \{ v^j \}_{j=1}^m$ otherwise, 
\begin{eqnarray*}
X_k &\stackrel{\rm def}{=}& \overbrace{X\otimes \cdots \otimes X}^{k-1}\otimes Z \otimes \overbrace{I\otimes \cdots \otimes I}^{\nu-k} \quad (1 \le k \le \nu)\nonumber \\
X_{k+\nu} &\stackrel{\rm def}{=}& \overbrace{X\otimes \cdots \otimes X}^{k-1}\otimes Y \otimes \overbrace{I\otimes \cdots \otimes I}^{\nu-i} \quad (1 \le k \le \nu)\\
X_{2\nu +1} &\stackrel{\rm def}{=}& \overbrace{X \otimes \cdots \otimes X}^{\nu}
\end{eqnarray*} 
and 
\begin{eqnarray*}
a_k &=& 0 \ (1 \le k \le \nu \ {\rm and} \ k=2\nu+1)\\
a_k  &=&   1 \ (\nu + 1\le k \le 2\nu).
\end{eqnarray*}
\end{Theorem}
That is, by choosing observables as above, $n$ copies of a standard $2$-dimensional maximally entangled state
gives a quantum correlation which is represented by a correlation function $\braket{u^i|v^j}$.
Although we used the fourth condition of Tsirelson's theorem (Theorem \ref{theorem: Tsirelson})
in the above theorem,
we can also use the third condition of the theorem. In this case, we just need to modify $\xi$ as
$\xi=d$ instead. 

\section{Optimality of the construction}
As we can easily see from the discussion in the previous section, by choosing observables appropriately, 
the state $\ket{\Phi_+}^{\otimes \nu}$ actually gives any quantum correlation function $c_{ij}=\left < u^i|v^j \right >$ satisfying
$\xi \le 2\nu +1$, where $\xi \stackrel{\rm def}{=} \min \left ( \dim {\rm span} \{ u^i \}_i^m,  \dim {\rm span} \{ v^j \}_j^n \right )$.
Moreover, when a quantum correlation function $c_{ij}$ is an extremal point of the set of all $m \times n$ quantum correlation functions, 
this choice of a state and observables is actually the unique optimal choice 
with respect to both the dimension of the
 Hilbert space and the amount of entanglement, up to local unitary equivalence.
This fact was also proved by Tsirelson \cite{Ts87a}. 
Before we state his theorem, we should note one important fact.
For an extremal $c_{ij}$, $\xi$ is uniquely determined, and called the rank of $c_{ij}$ \cite{Ts87a}.
That is, $c_{ij}=\left < u^i|v^j \right >= \left < u'^i|v'^j \right >$ implies $\xi=\xi'$ in this case.
Here, we just give the theorem, 
when the rank $\xi$ is even. The proof is found in \cite{Ts87a}.
\begin{Theorem}(Tsirelson)
Let $c_{ij}$ be an extremal point, with even rank $\xi$, 
of the set of all $m \times n$ quantum correlation functions.
Suppose a state $\rho$ and observables $\{ A_{i} \}_{i=1}^m, \{ B_j \}_{j=1}^n$  
on $\Hi_{A} \otimes \Hi_B$ 
satisfy Eq.(\ref{eq: Tsirelson's theorem 4}). 
Then, $\Hi _A$ and $\Hi_B$ can be decomposed as $\Hi _A= \bigoplus _{\alpha=1}^{L+1} \Hi_{A_{\alpha}}$
and $\Hi _B= \bigoplus _{\beta=1}^{L+1} \Hi_{B_{\beta}}$ such that $\dim \Hi_{A_{\alpha}} = \dim \Hi _{B_{\beta}} = 2^{\xi/2}$ for $1 \le \alpha \le L$
and $1 \le \beta \le L$. 
In this decomposition of the space, $A_i$ and $B_j$ can be written as 
\begin{eqnarray}\label{non-optimal observables}
A_i &=& \left ( \oplus _{\alpha =1}^{L} A_i^{(\alpha)} \right ) \oplus I_{L+1} \nonumber \\
B_j &=& \left ( \oplus _{\beta =1}^{L} B_j^{(\beta )} \right ) \oplus I_{L+1},
\end{eqnarray}
and $\rho $ can be written as
\begin{equation}\label{non-optimal state}
\rho  = \sum _{\alpha=1}^L \lambda _{\alpha} \ket{\Phi_{\alpha}}\bra{\Phi_{\alpha}}, 
\end{equation}
where $\{ \lambda _{\alpha } \}_{\alpha =1}^{L}$ is a probability distribution, and 
$\ket{\Phi_{\alpha}}\bra{\Phi_{\alpha}}$ is supported by $\Hi_{A_{\alpha}} \otimes \Hi_{B_{\alpha}}$.
Moreover, the state $\ket{\Phi_{\alpha}}$ and observables $\{ A_i^{\alpha} \}_{i=1}^m$ and 
$\{ B_j^{\alpha} \}_{j=1}^m$ on $\Hi_{A_{\alpha}} \otimes \Hi_{B_{\alpha}}$
is local unitary equivalent to the state $\ket{\Phi}^{\otimes \xi/2}$ and observables $\{ A_i \}_{i=1}^m$ and 
$\{ B_j \}_{j=1}^m$ defined in Theorem \ref{thm2} which corresponds to the same quantum correlation function $c_{ij}$.
\end{Theorem}  
Thus, the construction given in Theorem \ref{thm2} is optimal with respect to the dimension of the Hilbert space. 
For odd $\xi$, the corresponding theorem is slightly more complicated. 
However, the construction given in Theorem \ref{thm2} is optimal also in this case \cite{Ts87a}. 
Here, we note one more important fact which we can observe immediately from the above theorem.
Since the observables $\{ A_i \}_{i=1}^m$ and $\{ B_j \}_{j=1}^n$ defined by Eq.(\ref{non-optimal observables}) obey the anti-commutation relation,
they are actually proportional to a complex representation of the 
generators of the Clifford algebra $Cl(\mathbb{R}^{\xi})$. 
Thus, a set of observables corresponding to an extremal quantum correlation function always
has the structure of a Clifford algebra. 
Therefore, the construction using an irreducible representation of the Clifford algebra is optimal.

We can easily calculate the value of entanglement measures for the state $\rho$ defined by Eq. (\ref{non-optimal state}).
(The reader unfamiliar with entanglement measures is referred to \cite{entanglement measures}.) 
First, since we can deterministically derive $\ket{\Phi_+}^{\otimes \xi/2}$ by applying projective measurement to 
distinguish $\{ \Hi_{A_{\alpha}} \}_{\alpha=1}^{L+1}$, we derive a lower bound on 
the value of the distillable entanglement of $E_D(\rho) \ge \xi/2$.
Second, since Eq. (\ref{non-optimal state}) gives a pure-states-decomposition of $\rho$, we derive an upper bound
on the entanglement formation $E_F(\rho) \le \xi/2$. Moreover, since $\rho$ defined by Eq.(\ref{non-optimal state}) is
maximally correlated, the entanglement formation $E_F(\rho)$ is additive for $\rho$ and it coincides with the entanglement cost $E_C(\rho)$.
Since any entanglement measure $E(\rho)$ satisfies $E_D(\rho) \le E(\rho) \le E_C(\rho)$,
we can conclude $E(\rho)=E(\ket{\Phi}^{\otimes \xi/2})=\xi/2$.
Therefore, even if we use a mixed state $\rho$, we cannot reduce the amount of entanglement which 
is needed to represent an extremal 
quantum correlation function $c_{ij}$, and we actually need $\xi/2$ singlets to make such quantum correlation.
In this sense, the construction given in Theorem \ref{thm2} is 
an optimal construction with respect to both the dimension of the Hilbert space and the amount of entanglement.

We conclude this section with the following remark.
In some cases, the quantum state constructed above may be too large for any conceivable practical implementation.
However, there is a way to approximate the given correlations using a state with
much smaller dimensions. This is by using the Johnson-Lindenstrauss Lemma, which we state
in the form given in \cite{DG99}: 
\begin{Lemma} [Johnson-Lindenstrauss]
For $\epsilon \in (0,1)$ and $N$ a positive integer, let $K$ be a positive
integer such that
\begin{equation}
K \ge 4 ( \epsilon ^ 2 /2 - \epsilon ^ 3 /3 ) ^ {-1} log~N.
\end{equation}
Then for any set $V$ of $N$ points in $\mathbb{R}^d$ there is a mapping
$f: \mathbb{R}^d \rightarrow \mathbb{R}^K$ such that for all $u,v \in V$,
\begin{equation}
(1- \epsilon )  \braket{u | v} \le \braket{f(u) | f(v)}  \le (1 + \epsilon ) \braket{u | v}.                           
\end{equation}
\end{Lemma}

To use this lemma, we set $N=m+n$, $d=min(m,n)$, and let $V$ be the union of the point sets 
$\{ x _k \} _{k=1}^m$ and $\{ y_l \}_{l=1}^n$ from Theorem \ref{thm2}.
The lemma tells us that we may replace $V$ by point sets in $\mathbb{R}^K$, which is
a logarithmic reduction in dimension if $m$ and $n$ are comparable.
Theorem \ref{thm2} applied to these lower dimensional point sets
gives a quantum setting where the number of qubits required is also reduced 
logarithmically, and for which the correlations can be made arbitrarily close
to those desired.


\section{The maximum violation of quantum correlation vectors}\label{MV}
As we mentioned in Section \ref{section: quantum correlations},
quantum correlation vectors are not characterized by Tsirelson's Theorem \ref{theorem: Tsirelson},
which gives necessary and sufficient conditions for quantum correlation functions.
We saw that the theorem gives necessary conditions for quantum correlation vectors
that can be combined with the no-signalling condition.
These necessary conditions enable us to bound
the maximum quantum violation of (general) Bell inequalities~\cite{AII06a}.
However, it is not clear
to what extent the  no-signalling conditions
improve the bound given by
Tsirelson's theorem.

We denote
the set of quantum correlation functions (resp. quantum correlation vectors)
by $M_{\textrm{QB}}(m,n)$ (resp. ${\mathcal Q}_{\textrm{Cut}}(m,n)$)
following the notation of \cite{AII06a}.
The condition 3 of Theorem \ref{theorem: Tsirelson} ($\pm 1$ version)
can be represented by the \emph{elliptope}~\cite[section 28.4]{DL97a},
which is well studied in combinatorial optimization.
The elliptope $\varepsilon(G)$ of a graph $G = (V, E)$ with $n = |V|$ nodes
is the convex body consisting of vectors $z = (z_{ij}) \in \Re^E$
such that there exist unit vectors $u^i, u^j \in \Re^n$
for each node $i,j \in V$ satisfying $z_{ij} = \braket{ u^i | u^j }$.
Let $K_{m,n}$ be the complete bipartite graph
with nodes $V_{m,n} = \{ A_1, ..., A_m, B_1, ... , B_n \}$
and edges $E_{m,n} = \{ A_i B_j~|~1 \leq i \leq m, 1 \leq j \leq n \}$.
The elliptope $\varepsilon(K_{m,n})$
is the set of the vectors in $[-1, +1]^{mn}$
satisfying condition 3 of Theorem \ref{theorem: Tsirelson}, that is:
\begin{equation}\label{eq:Tsirelson}
M_{\textrm{QB}}(m,n) = \varepsilon(K_{m,n}).
\end{equation}
We can apply condition 3 of Theorem \ref{theorem: Tsirelson}
to quantum correlation vectors as follows:
\begin{itemize}
\item
For a quantum correlation vector $x = (c_{kl}) \in [-1, +1]^{m+n+mn}$,
there exist three sets of unit vectors
$\{ t^h \}_{h=1}, \{ u^i \}_{i=1}^m, \{ v^j \}_{j=1}^n \in \mathbb{R}^{d}$,
and $d \le 1+m+n$,
for which
\begin{eqnarray*}
 x  &= & (c_{kl}) \\ &= &
 ( \braket{ t^1 | u^i }_{1 \leq i \leq m},
\braket{ t^1 | v^j }_{1 \leq j \leq n}, \\
&\quad & \quad 
\braket{ u^i | v^j }_{1 \leq i \leq m, 1 \leq j \leq n} ).
\end{eqnarray*}
\end{itemize}
The set of the vectors satisfying the latter conditions corresponds to
the elliptope of another graph $\nabla K_{m,n}$ called the \emph{suspension graph} of $K_{m,n}$.
The graph $\nabla K_{m,n} = (\nabla V_{m,n}, \nabla E_{m,n} )$ is defined as the graph
with $1 + m + n$ nodes $\nabla V_{m,n} = \{ X \} \cup V_{m,n}$ and $m + n + mn$ edges
$\nabla E_{m,n} = \{ X A_i~|~ 1 \leq i \leq m \} \cup \{ X B_j~|~ 1 \leq j \leq n \} \cup E_{m,n}$.
The no-signalling conditions
correspond to the rooted semimetric polytope denoted by $\textrm{RMet}(\nabla K_{m,n})$,
which has been also well studied~\cite[section 27.2]{DL97a},
and is given by the following inequalities in the $\pm 1$ version:
\begin{eqnarray*}
x_{XA_i} + x_{XB_j} + x_{A_iB_j} & \ge & -1,\\
x_{XA_i} + x_{XB_j} - x_{A_iB_j} & \le & 1,\\
x_{XA_i} - x_{XB_j} + x_{A_iB_j} & \le & 1,\\
-x_{XA_i} + x_{XB_j} + x_{A_iB_j} & \le & 1.
\label{rsm2}
\end{eqnarray*}
Quantum correlation vectors are bounded as follows~\cite[Theorem 6]{AII06a}:
\begin{equation}\label{eq:upperbound}
{\mathcal Q}_{\textrm{Cut}}(m,n) \subseteq \varepsilon(\nabla K_{m,n}) \cap \textrm{RMet}(\nabla K_{m,n}).
\end{equation}

It has recently been shown that the inclusion in equation (\ref{eq:upperbound})
is in fact proper. Navascu\'es, et al. \cite{NPA} 
recently succeeded in
completely characterizing
the set of all quantum correlation vectors by means of a sequence of SDPs \cite{NPA}.
Using this characterization they got a tight upper bound on
the quantum violation of the Froissart Bell inequality, $I_{3322}$
(see \cite{Fro-NC81},\cite{CG04a}) of .25089.
This is significantly tighter than the bound of 0.3660 given by the right hand side
of (\ref{eq:upperbound}) for this inequality \cite{AII06a}.
A similar computational result was reported by Doherty et al. \cite{DLTW},
also using a hierarchy of SDPs.
The lower bound on the quantum violation is .25 \cite{CG04a}.

We begin by reviewing the work of  Navascu\'es, et al.
First, they treat a more general problem setting. That is,
their result is valid for all quantum correlation experiments which consist of measurements having 
any number of outcomes. 
So, we need to generalize the definition of a quantum correlation vector.
Suppose Alice's and Bob's measurements are described by projective measurements 
$\{ E_{\alpha}^{(i)} \}_{\alpha=1}^{a(i)}$ and $\{ F_{\beta}^{(j)} \}_{\beta=1}^{b(j)}$,
where $a(i)$ and $b(j)$ are the number of outcomes of the
$i$th and $j$th measurements of Alice and Bob, respectively. 
That is, this set of operators satisfies
\begin{eqnarray}\label{condition for projections}
&\ & E_{\alpha}^{(i)}=E_{\alpha}^{(i)\dagger}, \ F_{\beta}^{(j)}=F_{\alpha}^{(j)\dagger}, \nonumber \\
&\ & \sum_{\alpha=1}^{a(i)} E_{\alpha}^{(i)} = I, \ \sum_{\beta=1}^{b(j)} F_{\beta}^{(j)} = I,\nonumber \\
&\ & E_{\alpha}^{(i)}E_{\alpha'}^{(i)}=\delta _{\alpha \alpha'}E_{\alpha}^{(i)}, \
F_{\beta}^{(j)}F_{\beta'}^{(j)}=\delta _{\beta \beta'}F_{\beta}^{(j)}, 
\\ &\ &  [E_{\alpha}^{(i)}, F_{\beta}^{(j)}]=0. \nonumber
\end{eqnarray}
Then, by defining $P(\alpha|i) \stackrel{\rm def}{=} \bra{\Psi}E_{\alpha}^{(i)}\ket{\Psi} $, 
$P(\beta|j) \stackrel{\rm def}{=}  \bra{\Psi}E_{\beta}^{(j)}\ket{\Psi}$, 
$P(\alpha, \beta|i,j) \stackrel{\rm def}{=} \bra{\Psi}E_{\alpha}^{(i)}E_{\beta}^{(j)}\ket{\Psi}$,
a quantum correlation experiment is completely described by
the vector 
\begin{equation}
\left ( \{ P(\alpha|i) \}_{\alpha,i}, \{ P(\beta|j) \}_{\beta,j}, \{ P(\alpha, \beta|i,j)\}_{\alpha,\beta,i,j}  \right ),
\end{equation}
for
$1 \le \alpha \le a(i)-1$, $1 \le i \le m$, $1 \le \beta \le b(j)-1$, $1 \le j \le n$.
The set of these  vectors 
is denoted
as ${\mathcal Q}(m,n,a(i),b(j))$. We can easily see that  ${\mathcal Q}(m,n,a(i),b(j))$ is isomorphic to ${\mathcal Q}_{\textrm{Cut}}(m,n)$
when $a(i)=b(j)=2$ for all $i$ and $j$.
Suppose $\Upsilon $ is the set of all projections $I$, $E_{\alpha}^{(i)}$ and $F_{\beta}^{(j)}$
{\it for all $(\alpha, \beta, i, j)$ except $(a(i), b(j), i, j)$}; that is, $\alpha$ runs from $1$ to $a(i)-1$ and $\beta$
runs from $1$ to $b(j)-1$ .
We represent a product of projections in $\Upsilon$  
as a sequence on $\Upsilon$ (e.g. $E_{\alpha}^{(i)} $, $E_{\alpha}^{(i)}E_{\alpha'}^{(i')}$, $E_{\alpha}^{(i)}E_{\alpha'}^{(i')}F_{\beta}^{(j)}$ 
are sequences on $\Upsilon$).
For a sequence $S$, we define the length of $S$ as the minimum number of projections by which we can write $S$;
as a convention, we define the length of $I$ as $0$.
Then, we define $\mathcal{S}_c$ as the set of all sequences whose length is no greater than $c$; of course, 
we have $\mathcal{S}_c \subseteq \mathcal{S}_{c+1}$.
Let $\mathcal{S}_c=\{ S_k \}_{k=1}^{|\mathcal{S}_c|}$, where $|\mathcal{S}_c|$ is a size of a set $\mathcal{S}_c$. 
Then, we define a finite set of {\it independent equalities} $\mathcal{F}(\mathcal{S}_c)$
as a set of equalities of the form 
\begin{equation}\label{definition of equalities}
\sum_{kl}\left ( F \right )_{kl}\bra{\Psi}S_kS_l\ket{\Psi}=g \left ( P \right )
\end{equation}
 which is satisfied by sequences in $\mathcal{S}_c$ {\it for an arbitrary choice of projections 
$\{ E_{\alpha}^{(i)} \}_{\alpha=1}^{a(i)}$ and $\{ F_{\beta}^{(j)} \}_{\beta=1}^{b(i)}$} satisfying Eq.(\ref{condition for projections}).
In Eq.(\ref{definition of equalities}), $g \left ( P \right )$
is an affine function of probabilities $P(\alpha,\beta|i,j)$ defined as 
\begin{equation}
g \left ( P \right ) \stackrel{\rm def}{=} (g)_0 + \sum_{\alpha,\beta,i,j}(g)_{\alpha,\beta,i,j}P(\alpha,\beta|i,j).
\end{equation}
$\mathcal{F}(\mathcal{S}_c)$ is not uniquely determined from this definition.
But,  for all $c$ , there actually exists an easy algorithm to find a set $\mathcal{F}(\mathcal{S}_c)$ satisfying the above condition \cite{Miguel}. 
Since each equality in $\mathcal{F}(\mathcal{S}_c)$ is determined by a $|\mathcal{S}_c| \times |\mathcal{S}_c|$ 
matrix $F$ and an affine function $k$ on $\mathbb{R}^{|\mathcal{S}_c|}$,
we write $(F,k) \in \mathcal{F}(\mathcal{S}_c)$ if $F$ and $k$ satisfies Eq.(\ref{definition of equalities}) in $\mathcal{F}(\mathcal{S}_c)$. 

Now, we are ready to present the main result of Navascu\'es et al. \cite{NPA}.
Suppose ${\mathcal Q}_c(m,n,a(i),b(j))$ is defined as a set of quantum correlation vectors satisfying 
the following condition: there exists a $|\mathcal{S}_c| \times |\mathcal{S}_c|$ positive semidefinite matrix
$\Gamma \ge 0$ satisfying 
\begin{equation}
\Tr \left ( F^{T} \Gamma \right )=g \left ( P \right )
\end{equation}
for all $(F,k) \in \mathcal{F}(\mathcal{S}_c)$.
We have ${\mathcal Q}_{c+1}(m,n,a(i),b(j)) \subseteq {\mathcal Q}_c(m,n,a(i),b(j))$
Their main result is that \\
${\mathcal Q}(m,n,a(i),b(j))$ is actually a limit of 
the sequence of the sets $ \{ {\mathcal Q}_c(m,n,a(i),b(j)) \}_{c=0}^{\infty}$, that is,
\begin{equation}\label{complete characterization}
{\mathcal Q}(m,n,a(i),b(j)) = \bigcap _{c=0}^{\infty} {\mathcal Q}_c(m,n,a(i),b(j)).
\end{equation}
Moreover, by the definition,  a quantum correlation vector $P$ is in ${\mathcal Q}_c(m,n,a(i),b(j))$,
if and only if 
\begin{eqnarray*}
&\ & \max \{ \lambda \ |  \Gamma - \lambda I \ge 0 , \\ 
&\ & \qquad \quad \Tr \left ( F^{T} \Gamma \right )=g \left ( P \right ), \forall (F,k) \in \mathcal{F}(\mathcal{S}_c) \} \ge 0.
\end{eqnarray*}
Since the above optimization problem is an SDP, we can check whether $P$ is in ${\mathcal Q}_c(m,n,a(i),b(j))$ by
an SDP of the size $|\mathcal{S}_c| \times |\mathcal{S}_c|$.
Thus, we can judge whether $P$ is in ${\mathcal Q}_c(m,n,a(i),b(j))$ by this infinite sequence of SDPs.
In \cite{NPA}, they also gave a condition 
such that if it is satisfied by $\Gamma$ corresponding to a positive $\lambda$,
then we can immediately conclude $P \in {\mathcal Q}(m,n,a(i),b(j))$,
without having to solve
an infinite sequence of SPDs. 

Navascu\'es et al.'s method can be applied to the
calculation of maximal violation of Bell inequalities.
That is, we can always derive an upper bound of the maximal violation by optimizing Bell inequalities over ${\mathcal Q}_c(m,n,a(i),b(j))$,
and this optimization is also SDP \cite{NPA}. 
Suppose we are interested in a Bell inequality $J(P) = \sum _{\alpha, \beta, i, j} V(\alpha,\beta,i,j)P(\alpha,\beta|i,j)$,
for some coefficients
$V(\alpha,\beta,i,j)$.
Then, the optimization of $J(P)$ over ${\mathcal Q}_c(m,n,a(i),b(j))$ reduces to the following SDP:
\begin{eqnarray}\label{sequence maximally violation 1}
&\quad & \max \{ J(P)| P \in {\mathcal Q}_c(m,n,a(i),b(j))\} \nonumber \\
&=&\max \{ \Tr (\zeta _c \Gamma) \ | \ \Gamma \ge 0, \nonumber \\
&\quad & \quad  {\rm and} \ \Tr \left ( F^{T} \Gamma \right )=g \left ( P \right ), \forall (F,k) \in \mathcal{F}(\mathcal{S}_c) 
 \}.
\end{eqnarray} 
In the above equation, the $|\mathcal{S}_c| \times |\mathcal{S}_c|$ matrix $\zeta _c$ is defined as follows:
Since each index of the matrix $\zeta _c$ is associated to a sequence in $\mathcal{S}_c$, we directly represent each index
by the corresponding sequence. 
Then, $\left ( \zeta _c \right )_{E_{\alpha}^{(i)}, F_{\beta}^{(j)}}= V(\alpha,\beta,i,j)/2$ and 
$\left ( \zeta _c \right )_{S,S'}=0$ for all other combination of sequences $(S,S') \in \mathcal{S}_c \times \mathcal{S}_c$.
By Eq.(\ref{complete characterization}), we can easily see that 
Eq.(\ref{sequence maximally violation 1}) actually converge to the maximal violation of $J(P)$.

Also very recently, Doherty et al. derived a sequence of SDPs which  converge to the maximal violation 
of a Bell inequality \cite{DLTW}. In this final part of this section, we review their result.
Suppose $\mathcal{P}$ is a collection of Hermitian polynomials defined as follows:
As variables we consider $\{ E_{\alpha}^{(i)} \}_{\alpha,i}$ and $\{ F_{\beta}^{(j)} \}_{\beta,j}$.
The number of variables is equal to the number of projections which appear 
in the quantum correlation experiment, and are labeled in the same way as the projections. 
We should note that $\{ E_{\alpha}^{(i)} \}_{\alpha,i}$ and $\{ F_{\beta}^{(j)} \}_{\beta,j}$
are non-commutative variables, and do not necessary satisfy condition Eq.(\ref{condition for projections}).
We define a set of polynomials $\mathcal{T}_1$, $\mathcal{T}_2$, $\mathcal{T}_3$, $\mathcal{T}_4$, $\mathcal{T}_5$
on non-commutative variables $\{ E_{\alpha}^{(i)} \}_{\alpha,i}$ and $\{ F_{\beta}^{(j)} \}_{\beta,j}$ as follows:
\begin{eqnarray}
\mathcal{T}_1 &=& \Bigl\{ i[E_{\alpha}^{(i)},  F_{\beta}^{(j)}] \Bigm| \ 1 \le \alpha \le a(i), \nonumber \\ 
&\quad & \quad 1 \le \beta \le b(j), 1 \le i \le m, 1 \le j \le n  \Bigr\} \nonumber \\
\mathcal{T}_2 &=& \left ( \bigcup_i \{ I - \sum _{\alpha} E_{\alpha}^{(i)}  \}\right )
\cup \left ( \bigcup_j \{ I - \sum _{\beta} F_{\beta}^{(j)}  \}\right ) \nonumber \\
\mathcal{T}_3 &=& \left ( \bigcup_{i,\alpha} \{ ( E_{\alpha}^{(i)})^2 -E_{\alpha}^{(i)}  \}\right )
\cup \left ( \bigcup_{j,\beta} \{ ( F_{\beta}^{(j)})^2 -F_{\beta}^{(j)}  \}\right ) \nonumber \\
\mathcal{T}_4 &=& \left \{ i[E_{\alpha}^{(i)}, E_{\alpha'}^{(i)}] \right \}_{\alpha \neq \alpha',i} \cup 
\ \left \{ j[F_{\beta}^{(j)}, F_{\beta'}^{(j)}] \right \}_{\beta \neq \beta' ,j} \nonumber \\
\mathcal{T}_5 &=& \left \{ E_{\alpha}^{(i)} E_{\alpha'}^{(i)} + E_{\alpha'}^{(i)} E_{\alpha}^{(i)} \right \}_{\alpha \neq \alpha',i} 
\nonumber \\
&\quad & \quad \cup 
\ \left \{ F_{\beta}^{(j)} F_{\beta'}^{(j)}+ F_{\beta'}^{(j)} F_{\beta}^{(j)}  \right \}_{\beta\neq \beta',j}.
\end{eqnarray} 
We define $\mathcal{T} = \mathcal{T}_1 \cup \mathcal{T}_2 \cup \mathcal{T}_3 \cup \mathcal{T}_4 \cup\mathcal{T}_5$,
and $\mathcal{P}=\mathcal{T} \cup - \mathcal{T}$. 
Then, for a Hermitian $\{ E_{\alpha}^{(i)} \}_{\alpha,i}$ and $\{ F_{\beta}^{(j)} \}_{\beta,j}$,
$P$ is actually a set of Hermitian polynomials.
We should note that values of all polynomial in $\mathcal{P}$
are zero for  $\{ E_{\alpha}^{(i)} \}_{\alpha,i}$ and $\{ F_{\beta}^{(j)} \}_{\beta,j}$
satisfying Eq.(\ref{condition for projections}).
Then, we define the convex cone $\mathcal{C}_{\mathcal{P},c}$ as follows: $q$ is in $\mathcal{C}_{\mathcal{P},c}$,
then $q$ is a polynomial on $\{ E_{\alpha}^{(i)} \}_{\alpha,i}$ and $\{ F_{\beta}^{(j)} \}_{\beta,j}$ whose order is no greater than $2c$,
and which has the form $q=\sum _i r_i^{\dagger}r_i+\sum_i \sum_j s_{ij}^{\dagger}p_is_{ij}$, 
where $p_i \in \mathcal{P}$, and $r_i$, $s_{ij}$ are arbitrary polynomials on $\{ E_{\alpha}^{(i)} \}_{\alpha,i}$ and $\{ F_{\beta}^{(j)} \}_{\beta,j}$. 
Suppose a polynomial $q_{\nu}$ on $\{ E_{\alpha}^{(i)} \}_{\alpha,i}$ and $\{ F_{\beta}^{(j)} \}_{\beta,j}$ is defined as
\begin{equation}
q_{\nu} = \nu I - \left (  \sum _{\alpha, \beta, i, j} V(\alpha,\beta,i,j)E_{\alpha}^{(i)} F_{\beta}^{(j)} \right ).
\end{equation}
Then, the optimization problems
\begin{equation}
w_c \stackrel{\rm def}{=} \min \{ \nu | q_{\nu} \in \mathcal{C}_{\mathcal{P},c} \}
\end{equation}
converge to the maximal violation of the Bell inequality $J(P)$ in the limit $c \rightarrow \infty$.
Moreover, surprisingly, $w_c$ can also be calculated by SDP \cite{DLTW}. 
So, this optimization also gives another way to calculate an upper bound of maximal violation of a Bell inequality
as close as we like by SDP until the limit of our computational power.

Finally, we note one important fact.
 Even though ${\mathcal Q}(m,n,a(i),b(j))$ can be characterized 
by a sequence of SDPs, this does not imply that  
${\mathcal Q}(m,n,a(i),b(j))$  can be characterized by a polynomial time algorithm.
For example, 
if the convergence of ${\mathcal Q}_c(m,n,a(i),b(j))$ to ${\mathcal Q}(m,n,a(i),b(j))$ is slow,
then Eq.(\ref{complete characterization}) does not guarantee the existence of such a polynomial algorithm.
Actually, we can see a similar situation in the characterization of entangled states.  
The set of all separable (non-entangled) states can be characterized by a sequence of SDPs.
However, the problem to check whether a given state is separable or not
is actually NP-hard \cite{Ioannou}.
It is an important and interesting open question to see whether or not a
similar result holds for quantum correlation vectors.
 

\section*{Acknowledgements}
We would like to thank Miguel Navascu\'es for explaining his recent result 
and other recent related developments on this topic.
We would also like to thank Tsuyoshi Ito for bringing this work to our attention,
and for other helpful discussions. 
DA was supported by a grant from NSERC, and from the JSPS.
SM was supported by KAKENHI.
MO was
supported by Special Coordination Funds for Promoting
Science and Technology, the EPSRC 
grant EP/C546237/1, and the European Union Integrated 
Project Qubit Applications (QAP).

\appendix
\section{Clifford algebra}\label{section Clifford}
In this appendix, we give several important properties of the Clifford algebra associated 
to a real linear space $V=\mathbb{R}^n$, 
which are used in this paper. All of the facts given in this appendix can be found in \cite{Sn97a,Weyl}.

The Clifford algebra $Cl(V)$ associated to a real linear space $V=\mathbb{R}^n$ is defined to be the quotient algebra $T(V)/\mathcal{I}(V)$,
where $T(V) \stackrel{\rm def}{=} \bigoplus _{r=0}^{\infty}V^{\otimes r}$ is the tensor of $V$, 
and $\mathcal{I}(V)$ is the ideal of $T(V)$ generated by all element of the form $v \otimes v - <v|v>1$ for $v \in V$.
Although this is the definition of Clifford algebra, the following well-known characterization is more convenient in this paper.
\begin{Theorem}
Let $e_1, \cdots, e_n$ be any orthonormal basis of a real linear space $V$,
then, $Cl(V)$ is generated by $e_1, \cdots, e_n$ subject to the relations
\begin{equation}\label{eq: anti-commutation relation}
e_ie_j+e_je_i= 2\delta _{ij}.
\end{equation} 
\end{Theorem}
Thus, a set of all elements in the form $e_{i_1}\cdots e_{i_k}$ with $i_1<\cdots < i_k $ is a basis of Clifford algebra $Cl(V)$, 
and the dimension of $Cl(V)$ is $2^n$.

 In this paper, we are interested in the ``{\it complex representation}'' of Clifford algebra $Cl(V)$,
which is a real algebra homomorphism from $Cl(V)$ onto an algebra of all linear operators $\B (\Hi)$ on a complex finite-dimensional Hilbert space $\Hi$. 
The following property of irreducible complex representations of $Cl(\mathbb{R}^n)$ is important for our paper.
\begin{Theorem}
For even $n$, up to equivalence of representations,  
$Cl(\mathbb{R}^n)$ has only one complex irreducible representation,
whose dimension is $2^{\frac{n}{2}}$, and which is faithful.
For odd $n$, up to equivalence of representations,  
$Cl(\mathbb{R}^n)$ has only two different complex irreducible representations,
whose dimensions are $2^{\frac{n-1}{2}}$, and which are not faithful.
\end{Theorem}
We can make these irreducible representations in the following way.
Suppose $n=2 \nu$ is even, and $\Hi = \mathbb{C}^2$ is a $\nu$-qubit Hilbert space.
Then, on $\Hi$, the Weyl-Brauer matrices are defined as
\begin{eqnarray}
X_i &\stackrel{\rm def}{=}& \overbrace{Z\otimes \cdots \otimes Z}^{i-1}\otimes X \otimes \overbrace{I\otimes \cdots \otimes I}^{\nu-i}, \nonumber \\
X_{i+\nu} &\stackrel{\rm def}{=}& - \overbrace{Z\otimes \cdots \otimes Z}^{i-1}\otimes Y \otimes \overbrace{I\otimes \cdots \otimes I}^{\nu-i}, 
\end{eqnarray} 
where $X$, $Y$ and $Z$ are the Pauli matrices, and an index $i$ satisfies $1 \le i \le \nu$.
We can easily check that these matrices satisfy the anti-commutation relation (\ref{eq: anti-commutation relation}).
Thus, they are a representation of the generators of $Cl(\mathbb{R}^n)$. 
Since they also generate the algebra of all linear operators $\B(\Hi)$ which has the same dimension of $Cl(\mathbb{R}^n)$, 
this representation is faithful and irreducible.  
For odd $n=2\nu+1$, we can choose an irreducible representation of the 
generators of the Clifford algebra $Cl(\mathbb{R}^n)$ on 
a $n$-qubit Hilbert space $\Hi$
as $\{ X_i \}_{i=1}^{2\nu +1}$ where $\{ X_i \}_{i=1}^{2\nu}$ are defined by Eq.(\ref{definition X_i}) and 
$X_{2\nu +1} \stackrel{\rm def}{=} \overbrace{Z \otimes \cdots \otimes Z}^{\nu}$.
Again, we can check that the anti-commutation relation (\ref{eq: anti-commutation relation}), and 
$\{ X_i \}_{i=1}^{2\nu +1}$ generates the algebra of all linear operators $\B(\Hi)$, which agrees 
with the irreducibility of this representation. 
However, this representation is not faithful, since now the dimension of this algebra is smaller than that of $Cl(\mathbb{R}^n)$.
Another inequivalent irreducible representation is derived by redefining $-X_i$ as $X_i$ for $1 \le i \le \nu$.


\section{Review of the binary representation of the stabilizer formalism}\label{section: binary representation}
In Section \ref{problem setting} of this paper, we use the
{\it binary representation} of the stabilizer formalism.
For readers who are not familiar with this topic, we give a small review of it.
All the detailed proofs of the statements that appear in the following part can be found in \cite{G97} and \cite{NDM04}.

First, we rewrite the Pauli matrices by using binary indices as $I=\sigma _{00}, X= \sigma _{01}, Y=\sigma _{11}, Z=\sigma _{10}$.
Then, this notation gives an encoding of the Pauli matrices into the binary field $\mathbb{Z}_2$ as
\begin{equation}
I \leftrightarrow 00, \ X  \leftrightarrow 01, \ Y \leftrightarrow 11, \  Z \leftrightarrow 10.
\end{equation} 
More generally,  we can encode the tensor products of Pauli matrices into a vector space on the binary field as:
\begin{eqnarray}\label{eq; encoding n}
&\quad & \sigma _{(\vec{u},\vec{v})} \stackrel{\rm def}{=}\sigma _{u_1 v_1} \otimes \cdots \otimes \sigma _{u_N v_N} \nonumber  \\
& \leftrightarrow & (u_1 \cdots u_N|v_1 \cdots v_N)^T,
\end{eqnarray} 
where $\vec{u} \stackrel{\rm def}{=} (u_1 \cdots u_N)$, $\vec{v} \stackrel{\rm def}{=} (v_1 \cdots v_N)$, and $T$ is transposition.
We can easily see this representation satisfies 
$\sigma _{(\vec{u},\vec{v})} \dot \sigma _{(\vec{u'},\vec{v'})} = \alpha \sigma _{(\vec{u}\oplus\vec{u'},\vec{v}\oplus\vec{v'})}$,
where $\alpha$ is a phase factor which ranges over $\{\pm 1, \pm i\}$ depending on $\vec{u},\vec{v}, \vec{u'},\vec{v'}$, 
and $\oplus$ is bitwise summation in the binary field.
Since the Pauli group on $n$-qubits $\mathcal{P}_n$ can be written as 
\begin{equation}\label{eq; Pauli group}
\mathcal{P}_n = \{ \alpha \sigma _{(\vec{u},\vec{v})}| \alpha \in \{ \pm 1, \pm i \}, \vec{u}, \vec{v} \in \mathbb{Z}_2^n \},
\end{equation}
Eq.(\ref{eq; encoding n}) gives a representation of the $n$-qubits Pauli group on 
the $2n$ dimensional vector space on binary fields $\mathbb{Z}_2^{2n}$;
this representation is not faithful since we omit the phase factor $\alpha$ in Eq.(\ref{eq; Pauli group}).
We should note that the correspondence (\ref{eq; encoding n}) can be also understood as 
giving the projective representation of the finite field $\mathbb{Z}_2^{2n}$ on $n$-qubits Hilbert space.
    
Suppose $P$ is a $2n \times 2n$ binary matrix defined as $P \stackrel{\rm def}{=} \left [ \begin{array}{cc} 0 & I \\ I & 0 \end{array} \right ]$,
where $I$ is an $n \times n$ identity matrix. 
Then,  ``{\it $\sigma_{(\vec{u},\vec{v})}$ commutes with $\sigma_{(\vec{u'},\vec{v'})}$, 
if and only if $(\vec{u}, \vec{v})P(\vec{u'}, \vec{v'})^T=0$}.'' Thus, commutativity of elements of the Pauli group is 
equivalent to
the orthogonality of the
corresponding binary vectors under the symplectic inner product; $(\vec{u}, \vec{v})P(\vec{u'}, \vec{v'})^T$
is called the symplectic inner product between $(\vec{u}, \vec{v})$ and $(\vec{u'}, \vec{v'})$.    
Further, suppose $S$ is a commutative subgroup of the Pauli group on an $N$ qubit system.
Then, we can easily see the following fact: $\{ X_i \}_{i=1}^{m}$ are $m$ independent generators of $S$ if and only if 
a set of vectors $\{ \vec{e_k} \}_{k=1}^{m}$ on $\mathbb{Z}_2^{2N}$ corresponding 
to $\{ X_i \}_{i=1}^m$ is an
orthogonal basis of a subspace $\mathfrak{S}$ of $\mathbb{Z}_2^{2N}$
corresponding to $S$ in terms of the symplectic inner product;
in other words, the generator matrix $E \stackrel{\rm def}{=} \left ( \vec{e_1}, \cdots , \vec{e_m} \right )$ satisfies
$E^TPE = 0$. 
In this case, $\dim \mathfrak{S} =m$ and 
all pairs of different vectors $\vec{e}$ and $\vec{f}$ on $\mathfrak{S}$ 
satisfy $\vec{e}^T P \vec{f}=0$.  
In particular, the stabilizer group $S$ of an $N$-qubit stabilizer state corresponds to an $N$ dimensional self-dual subspace $\mathfrak{S}$;
that is, $\mathfrak{S}$ is equal to its symplectic orthogonal complement 
$\mathfrak{S}^c \stackrel{\rm def}{=} \{ \vec{e} \in \mathbb{Z}_2^{2N} |  \vec{e}^T P \vec{f} =0 \ (\forall \vec{f} \in \mathfrak{S})  \}$. 

Although a stabilizer group of a stabilizer state determines a corresponding self-dual subspace, 
a given $N$-dimensional self-dual subspace of a binary field, or a given generator matrix does not uniquely determine  
a corresponding stabilizer group and a stabilizer state. This fact can be seen as follows:
Suppose $\{ M_i \}_{i=1}^N$ is a set of commutative and independent Pauli operators on $N$-qubit Hilbert space,
and $E$ is a generator matrix corresponding to $\{ M_i \}_{i=1}^N$. Then, for all $(c_1, \cdots, c_n) \in \mathbb{Z}_N$,
$\{ (-1)^{c_i}M_i \}_{i=1}^N$ has the same matrix $E$ as its corresponding generator matrix. 
In this case, although a stabilizer state $\ket{\Psi}$ corresponding to $\{ (-1)^{c_i}M_i \}_{i=1}^N$ is not a stabilizer state corresponding to $\{ M_i \}_{i=1}^N$,
$\ket{\Psi}$ is a simultaneous eigenvector of $\{ M_i \}_{i=1}^N$.

Suppose $\{ \vec{e}_k \}_{k=1}^N$ is an orthogonal basis of a self dual subspace $\mathfrak{S}_N$,
and suppose $R$ is an $N \times N$ invertible binary matrix. 
Then, a set of vectors $\{ \vec{f_l} \}_{l=1}^N$ defined by 
\begin{equation}\label{def: f}
\vec{f_l} \stackrel{\rm def}{=} \sum _{k=1}^N \vec{e_k}R_{kl}
\end{equation}
is another orthogonal basis of $\mathfrak{S}_N$.
Suppose $\{ X_k \}_{k=1}^N$ are  independent generators of a stabilizer group $S_N$, 
$\{ \vec{e_k} \}$ are their binary representation, and $R$ is  an $N \times N$ invertible binary matrix. 
Then, $\{ a_k Y_k \}_{k=1}^N$ is another set of independent generators of $S_N$,
where $a_k \in \left \{\pm 1, \pm i \right \}$ is a suitably chosen phase factor, 
and $Y_k$ is an element of the Pauli group corresponding to a vector $\vec{f_k}$ defined by Eq.(\ref{def: f}).
Note that if $Y_k$ corresponds to the binary vector $\vec{f_k}$, 
then, $\pm Y_k$ and $\pm i Y_k$ also correspond to $\vec{f_k}$.

The Clifford group $\mathcal{G}_N$ is defined as the set of all unitary operators satisfying $U\mathcal{P}_N U^{\dagger} \subset \mathcal{P}_N$
for the $N$ qubits Pauli group $\mathcal{P}_N$.
We can easily see that an element $U$ of Clifford group $\mathcal{G}_N$ induces the linear transformation 
$f_U:\mathbb{Z}_2^{2N} \rightarrow \mathbb{Z}_2^{2N}$. 
In this case, it is known that the set of such transformations 
$\{ f_U \}_{U \in \mathcal{G}_N}$ coincides with the set of all symplectic transformation on $\mathbb{Z}_2^{2N}$.
Thus, for an element of the Clifford group $U$, $U \sigma _{\vec{e}}U^{\dagger}= a \sigma_{\vec{e'}}$ 
for a suitably chosen $a \in \{ \pm 1, \pm i\}$,
if and only if there exists a symplectic matrix $Q_U$ (a matrix satisfying $Q_U^TPQ_U=P$) such that $f_U(\vec{e})=Q_U \vec{e}=\vec{e'}$.


\section{Computational investigation of the strength of the no-signalling inequalities}\label{section: maximum violation}
We may wonder to what extent
the inclusion (\ref{eq:upperbound}) is essentially stronger
than Tsirelson's construction. 
In this appendix  we review an investigation of  a set of 
known Bell inequalities computationally,
showing that in most cases a tighter bound on quantum violation is achieved
by using the no-signalling conditions. In some cases, however, the bounds
are identical.

Takahashi et al. \cite{TMI} report on 
a
computational investigation of the optimal value of Bell inequalities
with the following constraints:
\begin{itemize}
\item[(i)] the elliptope $\varepsilon(\nabla K_{m,n})$
\item[(ii)] the elliptope $\varepsilon(\nabla K_{m,n})$
and the rooted semimetric polytope $\textrm{RMet}(\nabla K_{m,n})$
\end{itemize}
The optimal values were calculated by the method of Avis, Imai and Ito~\cite{AII06a}
using the software SDPA~\cite{FKNY04}
for a set of 89 Bell inequalities~\cite{IIA06},
which includes two quantum correlation functions:
A2 (the CHSH inequality) and A8 (the $I_{3322}$ inequality~\cite{CG04a}),
and the other 87 quantum correlation vectors.
The results are reproduced here as Table 1.
\begin{table}[t]
 \caption{The values of optimum solutions of the semidefinite programming
problems for those 89 Bell inequalities~\cite{IIA06}}
 \label{table:89}
 \begin{center}
  \setlength{\tabcolsep}{3pt}
  \footnotesize
  \begin{tabular}{|c|c|c|}\hline & (i) & (ii) \\\hline
\textbf{A1}  & 0.250 & 0.000 \\
\textbf{A2}  & \textbf{0.414} & \textbf{0.414} \\
\textbf{A3}  & 0.750 & 0.732 \\
\textbf{A4}  & 0.969 & 0.817 \\
\textbf{A5}  & 1.245 & 1.056 \\
\textbf{A6}  & 1.348 & 1.027 \\
\textbf{A7}  & 1.339 & 1.078 \\
\textbf{A8}  & \textbf{1.183} & \textbf{1.183} \\
\textbf{A9}  & 1.446 & 1.208 \\
\textbf{A10} & 1.444 & 1.106 \\
\textbf{A11} & 1.442 & 1.208 \\
\textbf{A12} & 1.463 & 1.245 \\
\textbf{A13} & 1.376 & 1.143 \\
\textbf{A14} & 1.471 & 1.283 \\
\textbf{A15} & 1.558 & 1.337 \\
\textbf{A16} & 1.554 & 1.252 \\
\textbf{A17} & 1.486 & 1.122 \\
\textbf{A18} & 1.484 & 1.144 \\
  \hline\end{tabular}
  \begin{tabular}{|c|c|c|}\hline & (i) & (ii) \\\hline
\textbf{A19} & \textbf{1.500} & \textbf{1.500} \\
\textbf{A20} & 1.703 & 1.560 \\
\textbf{A21} & 1.250 & 1.167 \\
\textbf{A22} & 1.669 & 1.516 \\
\textbf{A23} & 1.643 & 1.283 \\
\textbf{A24} & 1.685 & 1.388 \\
\textbf{A25} & 1.599 & 1.380 \\
\textbf{A26} & 1.660 & 1.280 \\
\textbf{A27} & 1.618 & 1.511 \\
\textbf{A28} & 1.590 & 1.527 \\
\textbf{A29} & 1.631 & 1.298 \\
\textbf{A30} & 1.730 & 1.243 \\
\textbf{A31} & 1.684 & 1.299 \\
\textbf{A32} & 1.662 & 1.161 \\
\textbf{A33} & 1.624 & 1.459 \\
\textbf{A34} & 1.545 & 1.410 \\
\textbf{A35} & 1.678 & 1.403 \\
\textbf{A36} & 1.662 & 1.224 \\
  \hline\end{tabular}
  \begin{tabular}{|c|c|c|}\hline & (i) & (ii) \\\hline
\textbf{A37} & 1.550 & 1.314 \\
\textbf{A38} & 1.584 & 1.273 \\
\textbf{A39} & 1.645 & 1.399 \\
\textbf{A40} & 1.676 & 1.392 \\
\textbf{A41} & 1.677 & 1.273 \\
\textbf{A42} & 1.667 & 1.431 \\
\textbf{A43} & 1.672 & 1.407 \\
\textbf{A44} & 1.570 & 1.437 \\
\textbf{A45} & 1.494 & 1.322 \\
\textbf{A46} & 1.570 & 1.255 \\
\textbf{A47} & 1.443 & 1.336 \\
\textbf{A48} & 1.414 & 1.294 \\
\textbf{A49} & 1.605 & 1.270 \\
\textbf{A50} & 1.579 & 1.410 \\
\textbf{A51} & 1.686 & 1.565 \\
\textbf{A52} & 1.665 & 1.501 \\
\textbf{A53} & 1.702 & 1.563 \\
\textbf{A54} & 1.690 & 1.489 \\
  \hline\end{tabular}
  \begin{tabular}{|c|c|c|}\hline & (i) & (ii) \\\hline
\textbf{A55} & 1.483 & 1.380 \\
\textbf{A56} & \textbf{1.717} & \textbf{1.717} \\
\textbf{A57} & 1.704 & 1.565 \\
\textbf{A58} & 1.769 & 1.564 \\
\textbf{A59} & 1.833 & 1.380 \\
\textbf{A60} & 1.482 & 1.161 \\
\textbf{A61} & 1.631 & 1.205 \\
\textbf{A62} & 1.528 & 1.199 \\
\textbf{A63} & 1.601 & 1.312 \\
\textbf{A64} & 1.644 & 1.229 \\
\textbf{A65} & 1.598 & 1.132 \\
\textbf{A66} & 1.610 & 1.360 \\
\textbf{A67} & 1.690 & 1.258 \\
\textbf{A68} & 1.647 & 1.276 \\
\textbf{A69} & 1.716 & 1.469 \\
\textbf{A70} & 1.829 & 1.612 \\
\textbf{A71} & 1.757 & 1.388 \\
\textbf{A72} & 2.103 & 1.784 \\
  \hline\end{tabular}
  \begin{tabular}{|c|c|c|}\hline & (i) & (ii) \\\hline
\textbf{A73} & 2.280 & 2.090 \\
\textbf{A74} & 2.060 & 1.830 \\
\textbf{A75} & 1.981 & 1.687 \\
\textbf{A76} & 1.733 & 1.420 \\
\textbf{A77} & 2.091 & 1.776 \\
\textbf{A78} & 2.193 & 2.106 \\
\textbf{A79} & 1.882 & 1.571 \\
\textbf{A80} & 1.486 & 1.220 \\
\textbf{A81} & 1.908 & 1.724 \\
\textbf{A82} & 1.629 & 1.369 \\
\textbf{A83} & 2.023 & 1.764 \\
\textbf{A84} & 1.972 & 1.791 \\
\textbf{A85} & 2.006 & 1.819 \\
\textbf{A86} & 2.075 & 1.850 \\
\textbf{A87} & 2.037 & 1.817 \\
\textbf{A88} & 1.584 & 1.370 \\
\textbf{A89} & 2.215 & 1.420 \\
&&\\
  \hline\end{tabular}
 \end{center}
\end{table}
As A2 and A8 are quantum correlation functions,
it is consistent with Tsirelson's Theorem \ref{theorem: Tsirelson}
that both optimal values are the same.
On the other hand, we also see that 
for two Bell inequalities, A19 and A56, 
both optimal values coincide.
This means that, for the Bell inequalities A19 and A56,
the rooted semimetric polytope $\textrm{RMet}(\nabla K_{m,n})$ is inactive.


\begin{thebibliography}{99}

\bibitem{AIIS05a}
D.~Avis, H.~Imai, T.~Ito, and Y.~Sasaki.
\newblock Two-party {Bell} inequalities derived from combinatorics via
  triangular elimination.
\newblock \emph{Journal of Physics A: Mathematical and General}, 38\penalty0
  (50):\penalty0 10971--10987, Nov. 2005.
\newblock arXiv:quant-ph/0505060.


\bibitem{AII06a}
D.~Avis, H.~Imai, and T.~Ito.
\newblock On the relationship between convex bodies related
to correlation experiments with dichotomic observables.
\newblock \emph{Journal of Physics A: Mathematical and General}, 39\penalty0
  (36):\penalty0 11283--11299, Sep. 2006.
\newblock arXiv:quant-ph/0605148.

\bibitem{AI07a}
D.~Avis and T.~Ito.
\newblock Comparison of two bounds of the quantum correlation set.
\newblock \emph{Proc. of ICQNM07},
  \penalty0  Guadeloupe, Jan. 2007.


\bibitem{Be64}
J.~Bell.
\newblock  On the Einstein-Podolsky-Rosen paradox.
\newblock \emph{Physics}, 1:195--200, 1964.

\bibitem{Ci80a}
B.~S. Cirelson.
\newblock Quantum generalizations of {Bell's} inequality.
\newblock \emph{Letters in Mathematical Physics}, 4\penalty0 (2):\penalty0
  93--100, 1980.

\bibitem{ClaHorShiHol-PRL69}
J.~F. Clauser, M.~A. Horne, A.~Shimony, and R.~A. Holt.
\newblock Proposed experiment to test local hidden-variable theories.
\newblock \emph{Physical Review Letters}, 23\penalty0 (15):\penalty0 880--884,
  Oct. 1969.

\bibitem{CHTW04a}
R.~Cleve, P.~H{\o}yer, B.~Toner, and J.~Watrous.
\newblock Consequences and limits of nonlocal strategies.
\newblock In \emph{Proceedings of 19th IEEE Annual Conference on Computational
  Complexity (CCC'04)}, pages 236--249, June 2004.
\newblock arXiv:quant-ph/0404076.

\bibitem{CG04a}
D.~Collins and N.~Gisin.
\newblock A relevant two qubit {Bell} inequality inequivalent to the {CHSH}
  inequality.
\newblock \emph{Journal of Physics A: Mathematical and General}, 37\penalty0
  (5):\penalty0 1775--1787, Feb. 2004.
\newblock arXiv:quant-ph/0306129.

\bibitem{DG99}
S.~Dasgupta and A.~Gupta. 
\newblock An elementary proof of the
Johnson-Lindenstrauss lemma. 
\newblock \emph{Technical Report TR-99-006}
International Computer Science Institute, Berkeley,
California, 1999.

\bibitem{DL97a}
M.~M. Deza and M.~Laurent.
\newblock \emph{Geometry of Cuts and Metrics}, 
\newblock Springer, May 1997.

\bibitem{DLTW}
A.~Doherty, Y-C.~Liang,~B. Toner,~S.,~Wehner.
\newblock The quantum moment problem and bounds on entangled multiprover games.
\newblock arXiv:0803.4373

\bibitem{experimental papers}S.J. Freedman and J.F. Clauser.
\newblock Experimental Test of Local Hidden-Variable Theories,
\newblock \emph{Physical Review Letters} 28:938-941, 1972;
A. Aspect, P. Grangier, and G. Roger.
\newblock Experimental Tests of Realistic Local Theories via Bell's Theorem,
\newblock \emph{Physical Review Letters} 47:460-463, 1981;
A. Aspect, P. Grangier, and G. Roger, 
\newblock Experimental realization of Einstein-Podolsky-Rosen-Bohm Gedankenexperiment: a new violation of Bell's inequalities. 
\newblock \emph{Physical Review Letters} 49:91-94, 1982;
Z. Y. Ou, and L. Mandel,  
\newblock Violation of Bell's inequality and classical probability in a two-photon correlation experiment. 
\newblock \emph{Physical Review Letters} 61;50-53, 1988. 

\bibitem{Fro-NC81}
M.~Froissart.
\newblock Constructive generalization of {Bell's} inequalities.
\newblock \emph{Il nuovo cimento}, 64B\penalty0 (2):\penalty0 241--251, 1981.

\bibitem{FKNY04}
K.~Fujisawa, M.~Kojima, K.~Nakata and M.~Yamashita.
\newblock SDPA(semidefinite programming algorithm) user's manual-version6.2.0, 2004.

\bibitem{G97}D. Gottesman, ``{\it Stabilizer Codes and Quantum Error Correction}'', PhD thesis, CalTech,
Pasadena, 1997.


\bibitem{JEB04}M. Hein, J. Eisert, and H. J. Briegel, 
\newblock Multiparty entanglement in graph states, 
\newblock {\it Physical Review A} 69, 062311, 2004

\bibitem{HDERNB06}M. Hein, W. Dur, J. Eisert, R. Raussendorf, M. Van den Nest, H.-J. Briegel, 
\newblock Entanglement in graph states and its applications, 
\newblock the Proceedings of the International School of Physics "Enrico Fermi" on 
"Quantum Computers, Algorithms and Chaos", Varenna, Italy, July, 2005, 
\newblock arXiv:quant-ph/0602096

\bibitem{Ioannou} L.M. Ioannou, 
\newblock Computational complexity of the quantum separability problem,
\newblock \emph{Quantum Information and Computation}, 7:335-370, 2007

\bibitem{IIA06}
T.~Ito, H.~Imai, and D.~Avis.
\newblock Bell inequalities stronger than the Clauser-Horne-Shimony-Holt inequality
for three-level isotropic states.
\newblock \emph{Physical Review A}, 73(042109), Apr. 2006.
\newblock arXiv:quant-ph/0605148.

\bibitem{Miguel} M. Navascu\'es, private communication.  

\bibitem{NPA}
M. Navascu\'es, S. Pironio, and A. Ac\'in.
\newblock A convergent set of semidefinite programs characterizing the set of quantum correlations.
\newblock arXiv:0803.4373

\bibitem{Pi91a}
I.~Pitowsky.
\newblock Correlation polytopes: Their geometry and complexity.
\newblock \emph{Mathematical Programming}, 50:\penalty0 395--414, 1991.

\bibitem{entanglement measures}
M.B. Plenio and S. Virmani.
\newblock An introduction to entanglement measures,
\newblock \emph{Quantum Information and Computation}, 7:1-51, 2007; 
R. Horodecki, P. Horodecki, M. Horodecki, and K. Horodecki,
\newblock Quantum entanglement,
\newblock e-print arXiv:quant-ph/0702225; J. Eisert and D. Gross, in
\newblock \emph{Lectures on Quantum Information}, edited by D. Bruss and G.
Leuchs, Wiley-VCH, Weinheim, 2006; M. Hayashi. 
\newblock \emph{Quantum Information, An Introduction}, Springer, Chapter 8, 2006. 

\bibitem{Sn97a}
J.~Snygg.
{\it Clifford Algebra}. 
Oxford University Press, 1997;
H. Blaine lawson, JR., Marie-Louise Michelsohn,
{\it Spin Geometry}.
Princeton University Press, 1989

\bibitem{TMI}
T.~Takahashi, S.~Moriyama, and H.~Imai.
\newblock A note on the upper bound derived by semidefinite 
programming for the maximum quantum violation of Bell inequalities.
\newblock AQIS 2008.

\bibitem{Tsi-HJS93}
B.~S. Tsirelson.
\newblock Some results and problems on quantum {Bell}-type inequalities.
\newblock \emph{Hadronic Journal Supplement}, 8\penalty0 (4):\penalty0
  329--345, 1993.

\bibitem{Ts87a}
B.~S. Tsirelson.
\newblock Quantum analogues of {Bell} inequalities: The case of two spatially
  separated domains.
\newblock \emph{Journal of Soviet Mathematics}, 36:\penalty0 557--570, 1987.

\bibitem{NDM04}M. Van den Nest, J. Dehaene, B. De Moor, 
\newblock Graphical description of the action of local Clifford transformations on graph states, 
\newblock {\it Physical Review A} {\bf 69}, 022316, 2004

\bibitem{WW01a}
R.~F. Werner and M.~M. Wolf.
\newblock All-multipartite {Bell}-correlation inequalities for two dichotomic
  observables per site.
\newblock \emph{Physical Review A}, 64\penalty0 (032112), Aug. 2001.
\newblock arXiv:quant-ph/0102024.


\bibitem{Weyl}H. Weyl, ``{\it The Classical Groups}'', Princeton University Press (1966);
R. Brauer, H. Weyl, 
\newblock Spinors in n Dimensions,
\newblock \emph{American Journal of Mathematics} 57:425-449 (1935)

\end{thebibliography}


\end{document}